\documentclass[sigconf]{acmart}

\usepackage{graphicx}
\usepackage{subcaption}
\copyrightyear{2026}
\acmYear{2026}

\setcopyright{cc}
\setcctype{by}

\acmConference[RecSys '26]
  {20th ACM Conference on Recommender Systems}
  {September 27--October 2, 2026}
  {Minneapolis, MN, USA}
\acmBooktitle{20th ACM Conference on Recommender Systems
  (RecSys '26), September 27--October 2, 2026,
  Minneapolis, MN, USA}
\acmDOI{10.1145/3773078.3831932}
\acmISBN{979-8-4007-2284-4/2026/09}

\begin{document}

\title{A Self-Triggered Agentic Push Recommendation System}


% \author{ByteDance Push Notification Team}
% \email{xxx@bytedance.com}
% \affiliation{%
%   \institution{ByteDance}
%   \city{Beijing}
%   \country{China}
% }

\author{Zhao-Yu Zhang}
\authornote{Zhao-Yu Zhang, Qingying Chen, and Chunyuan Zheng contributed equally.}
\email{zhaoyu.1@bytedance.com}
\affiliation{%
  \institution{ByteDance}
  \city{Beijing}
  \country{China}
}

\author{Qingying Chen}
\authornotemark[1]
\email{chenqingying@bytedance.com}
\affiliation{%
  \institution{ByteDance}
  \city{Beijing}
  \country{China}
}

\author{Chunyuan Zheng}
\authornotemark[1]
\email{zhengchunyuan@bytedance.com}
\affiliation{%
  \institution{ByteDance, Peking University}
  \city{Beijing}
  \country{China}
}

\author{Jing Zhou}
\email{zhoujing.23@bytedance.com}
\affiliation{%
  \institution{ByteDance}
  \city{Beijing}
  \country{China}
}

\author{Jian Sun}
\email{sunjian.2022@bytedance.com}
\affiliation{%
  \institution{ByteDance}
  \city{Beijing}
  \country{China}
}

\author{Siqi Chen}
\email{chensiqi.2025@bytedance.com}
\affiliation{%
  \institution{ByteDance}
  \city{Beijing}
  \country{China}
}

\author{Leiying Chen}
\email{chenleiying@bytedance.com}
\affiliation{%
  \institution{ByteDance}
  \city{Beijing}
  \country{China}
}

\author{Chuan Zhou}
\email{chuanzhou@bytedance.com}
\affiliation{%
  \institution{ByteDance}
  \city{Beijing}
  \country{China}
}

% \author{Xiang Li}
% \email{lixiang.2222@bytedance.com}
% \affiliation{%
%   \institution{ByteDance, Peking University}
%   \city{Beijing}
%   \country{China}
% }

\author{Huiyou Jiang}
\authornote{Huiyou Jiang is the corresponding author.}
\email{jianghuiyou.jhy@bytedance.com}
\affiliation{%
  \institution{ByteDance}
  \city{Beijing}
  \country{China}
}

\author{Xin Tao}
\email{taoxin.xt@bytedance.com}
\affiliation{%
  \institution{ByteDance}
  \city{Beijing}
  \country{China}
}

\author{Haoxuan Li}
\email{hxli@pku.edu.cn}
\affiliation{%
  \institution{Peking University}
  \city{Beijing}
  \country{China}
}

\author{Zhouchen Lin}
\email{zlin@pku.edu.cn}
\affiliation{%
  \institution{Peking University}
  \city{Beijing}
  \country{China}
}

\renewcommand{\shortauthors}{Zhao-Yu Zhang et al.}

\begin{abstract}
Push notification is a critical recommendation scenario on large-scale platforms, allowing the system to proactively reach users outside the application to improve long-term re-engagement. However, designing an optimal push system requires handling a complex action space for the \textbf{"whether and when"} delivery problem under strict system resource constraints. Existing solutions typically fall into two passive paradigms: \textit{pre-planned frequency} methods that allocate delivery times via offline modeling, limiting real-time adaptability; and \textit{fixed-interval triggering} methods that periodically poll the system, creating a strict dilemma between excessive computational overhead and diminished optimal timing capture. Furthermore, such multi-stage frameworks severely suffer from local optima. To overcome these limitations, in this paper, we propose \textbf{STEPS}, a proactive, \textbf{S}elf-\textbf{T}riggered \textbf{E}nd-to-end Agentic \textbf{P}ush Recommendation \textbf{S}ystem, which is already fully deployed at Douyin with over 1 billion users. STEPS reformulates push recommendation as a self-triggered agentic process in which the system decides not only whether to send a push, but also when to invoke itself again, thereby forming a closed loop that balances real-time effectiveness and efficiency. Specifically, STEPS consists of two decision transformer-based agents: a planning agent that schedules the next system invocation using a gated ordinal regression method, and an execution agent that decides whether to send a push based on trajectory rewards. Furthermore, we introduce a lightweight filtering agent to both control computational overhead and act as a crucial safeguard against unreasonable planning behaviors. Online A/B testing demonstrates that STEPS significantly increases user active days by \textbf{0.2843\%} and reduces the push permission disablement rate by \textbf{1.9089\%}, while the filtering agent reduces computational overhead by \textbf{79.42\%}.

\end{abstract}

% what不写
% 不trivial，带time排序，不能每时每秒ranking
% base方法更易懂一些



\maketitle

\begin{figure}[t]
\centering
% \vspace{-5pt}
{
\begin{minipage}[t]{\linewidth}
\centering
\includegraphics[width=0.85\textwidth]{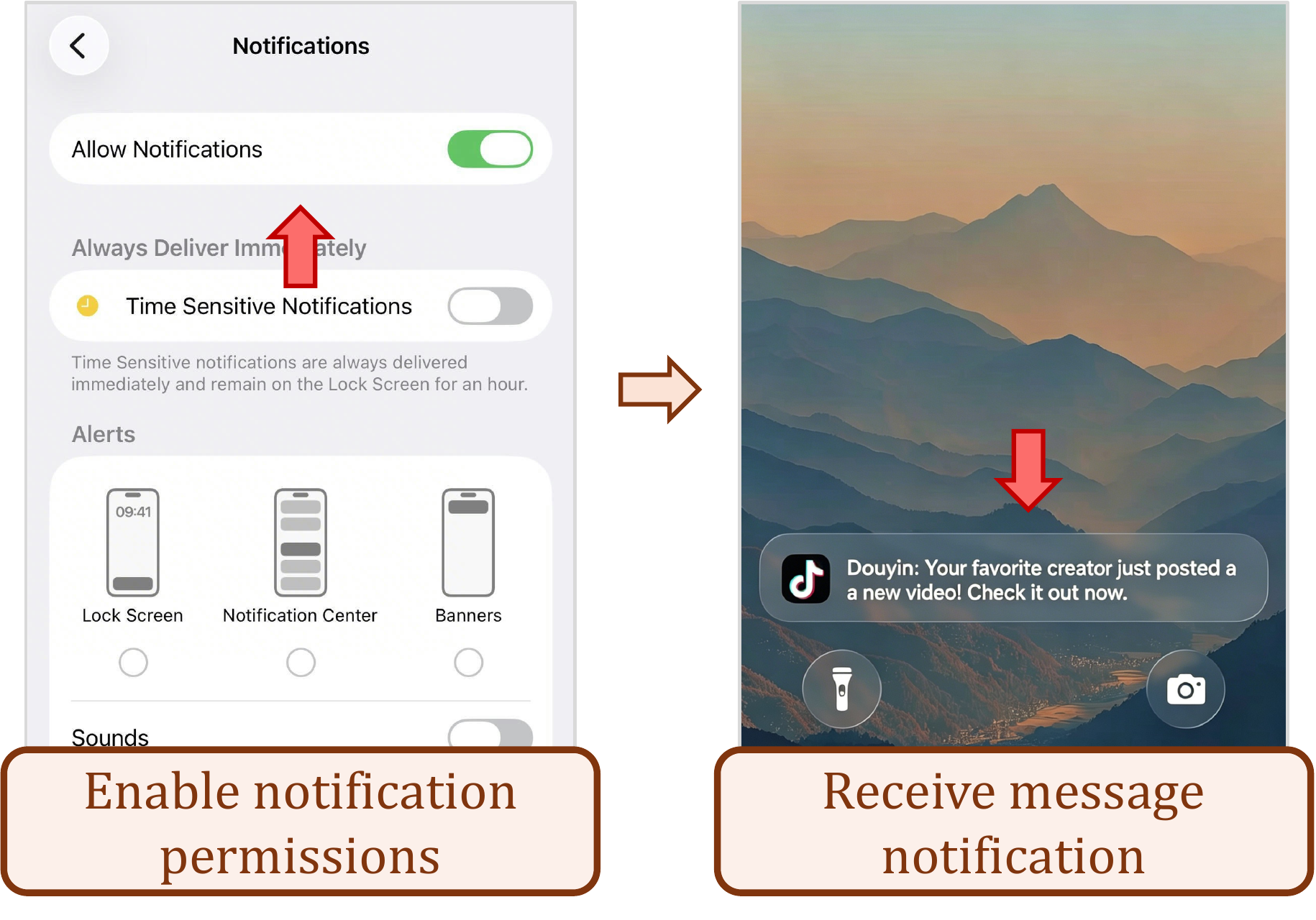}
\end{minipage}%
}%
\\
\centering
\caption{Push notification examples.}
% \vspace{-8pt}
\label{fig:motivation}
\end{figure}

\section{Introduction}
Push notification is an important recommendation channel because it enables the system to proactively reach users outside the application, rather than waiting for them to launch the application on their own~\cite{tang2013design,pielot2018dismissed,wohllebe2020consumer}. {As shown in Figure 1, when users enable the push permission (left), they will receive the notification (right)}. At the same time, push notification is also a challenging problem: the system must decide whether and when to deliver a notification~\cite{pham2016effects,wohllebe2021mobile}. Well-designed pushes can improve re-engagement and user activity, but poorly timed ones lead to push notification disablement because of user fatigue~\cite{stroud2020effects,wheatley2021temporal,barnes2025push}.

% Solving the "whether and when" problem is highly non-trivial, since push delivery is a continuous and sequenced optimization problem~\cite{mumcu2025you,abboud2025agentic}. To find the optimal delivery moment, a naive approach would be to evaluate every user every second. However, in large-scale industrial systems, this per-second real-time ranking is impractical due to system resource constraints. Existing solutions can be divided into two main paradigms. The first is a two-stage approach that relies on offline user-level uplift modeling to calculate the positive and negative value of each possible timing and frequency, and then uses integer programming solvers to solve optimal times and frequencies~\cite{gupta2016email,yancey2020sleeping}.  The second periodically activates push systems based on a pre-defined short time interval to decide whether to send a push~\cite{lee2016assorted,zhao2018notification,aharon2019soft}.

Solving the "whether and when" problem is highly non-trivial, since push delivery is a continuous and sequenced optimization problem~\cite{mumcu2025you,abboud2025agentic}. To find the optimal delivery moment, a naive approach would be to evaluate every user every second. However, in large-scale industrial systems, this per-second real-time ranking is impractical due to system resource constraints. Existing solutions can be divided into two main paradigms. The first paradigm, \textit{pre-planned frequency} methods, is a two-stage approach that relies on offline user-level uplift modeling to calculate the positive and negative value of each possible timing and frequency, and then uses integer programming solvers to allocate optimal times and frequencies~\cite{gupta2016email,yancey2020sleeping}. The second paradigm, \textit{fixed-interval triggering} methods, periodically activates push systems based on pre-defined short time intervals to decide whether to send a push~\cite{lee2016assorted,zhao2018notification,aharon2019soft}.

% However, these methods have several limitations. First, making a decision based on a pre-defined time points and frequencies without considering real-time information, such as most recent user activity and push content, will lead to unreasonable pushes. For example, a push may not be sent even when the trajectory suggests that a current push could be valuable. In addition, activating the system periodically will result in too many execution processes, consuming a lot of resources since the item sorting involves heavy computation. Finally, the multi-stage framework may find the local optima of each stage, leading to an overall suboptimal performance.
However, these methods have several limitations. First, pre-planned frequency methods make decisions based on pre-defined time points without considering real-time information, such as recent user activity and push content. For example, a push may not be sent even when real-time trajectories suggest it would be highly valuable. Second, fixed-interval triggering methods face a strict dilemma: activating the system too frequently consumes excessive computational resources due to heavy item-sorting workloads, while broader intervals fail to capture the optimal delivery moments. Finally, multi-stage frameworks often fall into local optima at each stage, leading to overall suboptimal performance.

% To address these issues, we propose \textbf{STEPS}, a proactive, \textbf{S}elf-\textbf{T}riggered end-to-end \textbf{A}gentic Push \textbf{R}ecommendation \textbf{S}ystem for push notification, which has the ability to execute actions and proactively generates the time for its own next decision. Specifically, the agentic system is composed of three agents: a planning agent, an execution agent, and a filtering agent. The overall pipeline works as follows. First, the planning agent provides the predicted next push time. Second, when the time arrives, the execution agent decides whether to send such a push to users. At the end of this loop, the planning agent updates a new push time. Finally, we introduce a lightweight filtering agent that screens out both low-value requests before they consume heavy computation and unreasonable planning behaviors to keep the push system safe. 
Recent advances in agentic AI have increasingly emphasized closed-loop systems
that plan and act based on feedback from their environments \cite{silver2025welcome,huang2025towards}. Motivated by this perspective, we
reformulate push recommendation as a self-triggered agentic process, proposing \textbf{STEPS}, a proactive, \textbf{S}elf-\textbf{T}riggered \textbf{E}nd-to-end Agentic \textbf{P}ush Recommendation \textbf{S}ystem, which has the ability to execute actions and proactively generate the time for its own next decision in a closed loop. In other words, the agentic system controls not only what action to take, but also when it will be invoked again. Specifically, the agentic system is composed of three agents: a planning agent, an execution agent, and a filtering agent. The overall pipeline operates in a closed loop: first, the planning agent predicts the optimal next push time. When this time arrives, the execution agent evaluates real-time context to decide whether to send a push. At the end of this step, the planning agent generates a new future time. Finally, we introduce a lightweight filtering agent that screens out both low-value requests before they consume heavy computation and unreasonable planning behaviors to ensure system safety.

\begin{figure*}[t]
\centering
% \vspace{-5pt}
{
\begin{minipage}[t]{\linewidth}
\centering
\includegraphics[width=0.85\textwidth]{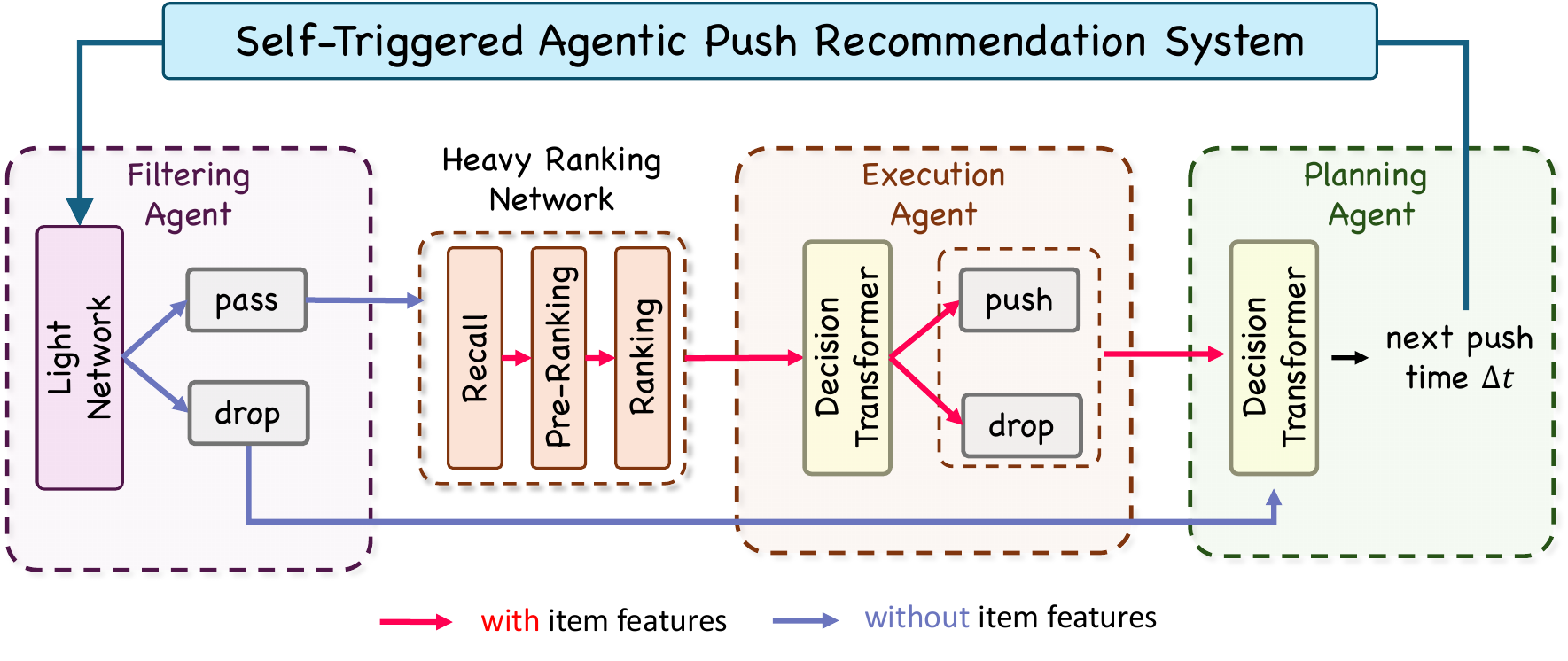}
\end{minipage}%
}%
\\
\centering
% \vspace{-4pt}
\caption{Overall framework.}
% \vspace{-4pt}
\label{fig:overall}
\end{figure*}

% 用户发起，system实时响应
% request level
% 具体decision

We summarize our main contributions as follows: 

\begin{itemize}
    \item We reformulate industrial push recommendation as a self-triggered closed-loop decision problem, and propose a novel agentic end-to-end push recommendation system (STEPS), which proactively generates the next push time (planning agent) and determine whether to push (execution agent), with a lightweight filtering agent to both control the computational overhead and ensure system safety.
    
    \item For push time generation, we propose a novel gated ordinal regression method to effectively inject target returns. For execution, we propose a value-guided learning method to correct suboptimal behaviors in offline logs.

    \item STEPS has been fully deployed on Douyin, which has over 1 billion users, showing that such a pipeline can both improve user active days and reduce the push permission disablement rate. In addition, we show the filtering agent can significantly reduce the computational overhead.
\end{itemize}

\section{Related Work}
\subsection{Push Notification Systems}
Push notification systems are multi-dimensional decision making systems that involve the content to send, and more importantly, \textit{whether} and \textit{when} to send. The problem of determining push content has become standard in industrial systems~\cite{zhao2017push,loni2019push,yue2022pairwise}, and in this paper we focus on the "whether and when" problem. To solve this problem, existing works generally fall into two paradigms. The first paradigm, which we refer to as \textbf{Pre-planned Frequency} methods, focuses more on the timing aspect. These methods usually conduct offline user-level uplift modeling and utilize dynamic programming algorithms to solve for the optimal push frequency and allocate delivery timings in advance. When the pre-planned time arrives, the system is then triggered to execute the sending plan~\cite{gupta2016email,yancey2020sleeping,satoh2019adaptive,yuan2022survival,li2026user}. However, scheduling push notifications offline in advance results in reduced timeliness and diminishes their real-time effectiveness. Meanwhile, the second paradigm, known as \textbf{Fixed-interval Triggering} methods, bypasses the timing allocation issue by periodically polling the system at fixed time intervals to make real-time judgments on \textit{whether} to send a notification~\cite{yuan2025generative_linkedin,bonner2018causal,saito2020unbiased,wang2021deconfounded,chen2024upliftrec}. However, time slices that are too broad result in lower-value push notifications, while slices that are too granular consume excessive computational resources. The limitations of these previous push paradigms call for a self-triggered
agentic process that jointly determines whether to send a push and when to invoke the system again, while preserving real-time performance and resource
efficiency.

\subsection{Generative Reinforcement Learning}
Reinforcement learning has been widely studied for optimizing long-term objectives in recommendation, including slate recommendation, feed ranking, and notification delivery~\cite{ie2019slateq,zou2019feedrec,yuan2022offline,prabhakar2022multiobjective}. However, deploying conventional value-based or policy-gradient methods in large-scale industrial systems can sometimes present challenges, as these approaches may be susceptible to unstable bootstrapping and sensitivity to distribution shifts in offline logs. To address this, recent generative RL methods offer an alternative by formulating sequential decision-making as a conditional sequence modeling problem. Decision transformer (DT)~\cite{chen2021decision} predicts actions conditioned on return-to-go, historical states and past actions through a supervised learning objective. Trajectory Transformer~\cite{janner2021trajectory} further treats states, actions, and rewards as a unified trajectory sequence for long-horizon planning, and Q-learning decision transformer~\cite{yamagata2023qdt} further improves trajectory stitching by relabeling returns with dynamic programming, thereby combining the stitching ability of Q-learning with the training stability of DT. This paradigm has also shown promise in industrial multi-objective systems, such as notification optimization~\cite{yuan2025generative_linkedin} and auto-bidding~\cite{gao2025generative_gave}. Our work follows this direction but targets proactive push recommendation, where the system must jointly generate \textit{when} to activate the next request and decide \textit{whether} to send after activation.

% \newpage
\section{Method}

% For a specific user at time $t$, we have the features $x_u^t$ captures user-side information such as device properties and historical activity, $x_c^t$ represents contextual information such as current time and current-day activity status, and $x_i^t$ denotes candidate content features.

\begin{figure*}[t]
\centering
% \vspace{-5pt}
{
\begin{minipage}[t]{\linewidth}
\centering
\includegraphics[width=1\textwidth]{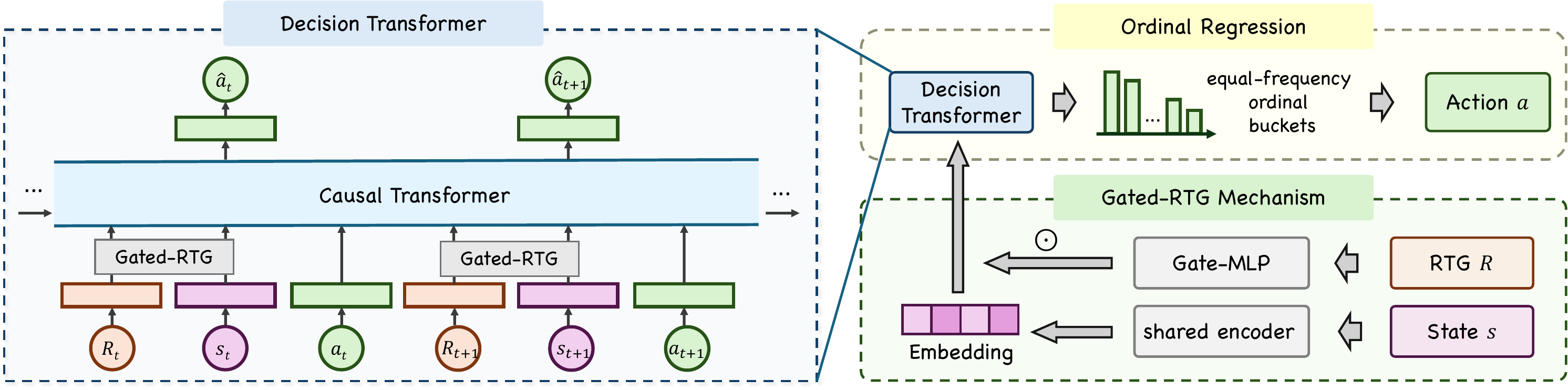}
\end{minipage}%
}%
\\
\centering
\caption{Architecture of planning agent.}
% \vspace{-8pt}
\label{fig:planing_agent}
\end{figure*}

\subsection{Problem Setup}
% Generally, optimizing a push recommendation system for long-term engagement can be formulated as a sequential execution problem over a continuous time horizon. Suppose there are $T$ timestamps to push within a time period (we use one day as an example), the goal is to maximize the positive value (e.g., daily active) while controlling the daily negative value (e.g., notification disabled) over the $T$ candidate push timestamps, where $T$ can be varied for different users and methods. For a specific user at each time $t \in \{1,2,\ldots, T\}$, unlike standard RL that optimizes a policy purely via reward maximization, recent advances in decision transformers (DT) formulate this as a conditional sequence modeling problem. Let $R_t$ denote the return-to-go (RTG) at step $t$, defined as the desired future return:
% \begin{equation*}
% R_t = \sum_{j=t}^{T} \left(r^+_j - \lambda_n r^-_j\right),
% \end{equation*}where $r^+$ and $r^-$ are positive and negative rewards respectively, and $\lambda_n$ is a hyperparameter balancing the trade-off. 
Generally, optimizing a push recommendation system for long-term engagement can be formulated as a sequential execution problem over a continuous time horizon. Suppose there are $T$ candidate timestamps for a push within a given time period (e.g., one day). The goal is to maximize the positive value (e.g., user active days) while controlling the daily negative value (e.g., notification disablement) over these $T$ timestamps, where $T$ can vary across different users and methods. For a specific user at each time $t \in \{1,2,\ldots, T\}$, unlike standard RL that optimizes a policy purely via reward maximization, recent advances in decision transformers (DT) formulate this as a conditional sequence modeling problem. Let $R_t$ denote the return-to-go (RTG) at step $t$, defined as the desired future return:
$$ R_t = \sum_{j=t}^{T} \left(r^+_j - \lambda_n r^-_j\right), $$
where $r^+$ and $r^-$ are positive and negative rewards respectively, and $\lambda_n$ is a hyperparameter balancing the trade-off.

% In previous DT-based recommendation frameworks, the model takes the current state $s_t$, including user feature, item feature, and environment information (such as current time and previous push time), and the target RTG $R_t$ as input, and conditionally generates the corresponding binary action $a_t \in \{0, 1\}$:
% \begin{equation*}
% a_t = \pi(s_t, R_t).
% \end{equation*}
In previous DT-based recommendation frameworks, the model takes the current state $s_t$, including user features, environment information (such as current time and previous push time), the target RTG $R_t$, and optional item features as input, and conditionally generates the corresponding binary action $a_t \in \{0, 1\}$:
$$ a_t = \pi(s_t, R_t). $$

% The first type of previous multi-stage methods pre-calculate $T$ offline before the day based on uplift modeling and integer optimization, and send a push at time $t \in \{1,2,\ldots,T\}$, which does not consider the timely environment information and may find local optima. The second type methods periodically activate the system based on a pre-defined short time interval, which consumes resources and involves many low‑value push timings.
As discussed, existing approaches struggle with this problem. The first paradigm, pre-planned frequency methods, pre-calculates the $T$ timestamps offline prior to the target day using uplift modeling and integer optimization. Consequently, they execute pushes at fixed times $t \in \{1,2,\ldots,T\}$ without considering real-time environment information, resulting in local optima. Alternatively, fixed-interval triggering methods periodically activate the system on pre-defined short time intervals, which consumes excessive computational resources and evaluates many low-value push timings.

% To bridge this gap, we formulate push recommendation as a sequential generate-and-decide problem, in which the system sequentially generates the next push time and decides whether to send a push based on the timely environment information, discarding the pre-defined $T$. Specifically, if there is a push at time point $t$, the planning agent will output $\Delta_t > 0$ based on $s_t$, which defines the time gap from now. Then the system will be activated at $t+\Delta_t$. Next, the system executes a binary action $a_{t+\Delta_t} \in \{0,1\}$ based on $s_{t+\Delta_t}$ and RTG $R_{t+\Delta_t}$ to determine whether to send the push. Finally, the planning agent outputs a next push time, which defines the time gap between now ($t+\Delta_{t}$) and the next push. Formally, under our formulation, we need to determine the action tuple ($a$, $\Delta$) simultaneously at time $t$ in an end-to-end manner:
% \begin{equation*}
% (\Delta_t, a_{t+\Delta_t}) = (\pi_P(s_t, R_t),\pi_A(s_{t+\Delta_t}, R_{t+\Delta_t})),
% \end{equation*}where $\pi_P$ and $\pi_A$ are the planning agent and execution agent, respectively. The overall structure of our framework is shown in Figure~\ref{fig:overall}. To ensure an end-to-end learning, we share the bottom user, environment, and item embedding parameters for planning and action tasks. 
To bridge this gap, we formulate push recommendation as a sequential generate-and-decide problem, completely discarding the reliance on a pre-defined $T$. Instead of a fixed schedule, the system dynamically generates its next activation time and makes push decisions based on real-time environment information. Specifically, whenever the system is activated at a timestamp $t$, it observes the current state $s_t$ and the target RTG $R_t$. It then simultaneously performs two tasks: the execution agent $\pi_A$ outputs a binary decision $a_t \in \{0,1\}$ to determine whether to send a push right now, and the planning agent $\pi_P$ outputs a time gap $\Delta_t > 0$ to schedule the next system activation at time $t + \Delta_t$. Formally, under our formulation, the system determines the action tuple $(a_t, \Delta_t)$ in an end-to-end manner:
$$ (a_t, \Delta_t) = (\pi_A(s_t, R_t), \pi_P(s_t, R_t)), $$
where $\pi_A$ and $\pi_P$ are the execution agent and planning agent, respectively. The overall structure of our framework is shown in Figure~\ref{fig:overall}. To enable end-to-end learning, we share the bottom user, environment, and optional item embedding parameters for both planning and action tasks.

% In addition, the reason for using DT as a backbone for $a_t$ and $\Delta_t$ is that the push notification environments are noisy, and DT is good at imitation and sequence modeling problems than RL, thus has the potential to build a robust framework to improve the business value.

% 这里是否需要画trigger gate进Figure 2

% 信号是
% 求解不用写，rule based 兜底

\subsection{Planning Agent}

% The planning agent acts as the proactive engine of our agentic framework. {As illustrated in Figure 3, the DT model structure requires both the state $s_t$ and a target RTG $R_t$ as inputs, and the generated $\Delta_t$ as outputs.} Note that the item side information is optional, since it depends on whether we do a item sorting. To effectively achieve this generative capability in a highly dynamic push scenario, we introduce two core methodologies:
The planning agent acts as the proactive engine of our agentic framework. As illustrated in Figure~\ref{fig:overall}, the DT model structure requires both the state $s_t$ and a target RTG $R_t$ as inputs, and conditionally outputs the generated $\Delta_t$. Note that the item side information is optional, as it depends on whether we perform item sorting. To effectively achieve this generative capability in a highly dynamic push scenario, we introduce two core mechanisms:

% \textbf{Gated-RTG Mechanism for Condition Injection.} The standard architectures that simply concatenate RTG with dense, high-dimensional state representations will cause the RTG to be easily overwhelmed by the dominant state features, which makes the model difficult to learn stable correlations. To address this, we propose a Gated-RTG structure. Instead of naive concatenation, the RTG $R_t$ is first passed through a Multi-Layer Perceptron (MLP) tower to upscale its dimensionality. Then, the RTG is injected into the state via an element-wise dot product (gating mechanism) by
% $e_{gated} = e_s \odot \text{MLP}(R_t)$, where $e_s$ is the state embedding. This gating mechanism ensures that the RTG is not overwhelmed by state features. The gated representation, $e_{gated}$, is then fed into the decoder to generate the push time.
\noindent
\textbf{Gated-RTG Mechanism for Condition Injection.} Standard architectures that simply concatenate the RTG with dense, high-dimensional state representations often cause the RTG to be easily overwhelmed by the dominant state features, making it difficult for the model to learn stable correlations. To address this, we propose a Gated-RTG structure. Instead of naive concatenation, the RTG $R_t$ is first passed through a Multi-Layer Perceptron (MLP) tower to upscale its dimensionality. Then, the RTG is injected into the state via element-wise multiplication (Hadamard product), acting as a gating mechanism: $e_{gated} = e_s \odot \text{MLP}(R_t)$, where $e_s$ is the state embedding. The gated representation, $e_{gated}$, is then fed into the decoder to generate the push time.

% \textbf{Ordinal Regression for Time Generation.} Predicting an exact continuous time gap $\Delta_t$ (ranging from 0 up to 24 hours) via standard regression is unstable, since the label distribution is non‑Gaussian, heavily long‑tailed, and the tolerance to errors is non‑linear. For example, the difference between $10$ minutes and $40$ minutes is far more critical than that between $10$ hours and $10.5$ hours. To overcome this, we reformulate the time generation step as an ordinal regression problem~\cite{burkner2019ordinal}, since there is an overall monotonic relationship between action and reward. For example, as the push gap increases, the active rate decreases. Specifically, we divide the continuous action space into $K=100$ equal-frequency buckets with boundaries $\{\tau_k\}_{k=1}^K$. The decoder outputs ordinal logits $z_{k}$, where each bucket represents the probability that the next push time $\Delta_t$ is strictly greater than the boundary $\tau_k$. Therefore, for a specific user, the label in $k$-th bucket is denoted as $y_{k} = \mathbf{1}[\Delta_t > \tau_k]$, the model is optimized using a Binary Cross-Entropy (BCE) loss across all buckets:
% $$\mathcal{L}_{next} = \sum_{k=1}^{K}BCE(y_{k}, z_{k}).$$
\noindent
\textbf{Ordinal Regression for Time Generation.} Predicting an exact continuous time gap $\Delta_t$ (ranging from 0 up to 24 hours) via standard regression is unstable, because the label distribution is non-Gaussian, heavily long-tailed, and the tolerance to errors is non-linear. For example, the difference between $10$ minutes and $40$ minutes is far more critical than that between $10$ hours and $10.5$ hours. To overcome this, we reformulate the time generation step as an ordinal regression problem~\cite{burkner2019ordinal}, leveraging the overall monotonic relationship between action and reward (e.g., as the push gap increases, the active rate typically decreases). Specifically, we divide the continuous action space into $K=100$ equal-mass buckets with boundaries $\{\tau_k\}_{k=1}^K$, ensuring each bucket contains the same number of samples. The decoder outputs ordinal logits $z_{k}$, where each bucket represents the probability that the next push time $\Delta_t$ is strictly greater than the boundary $\tau_k$. Therefore, for a specific user, the label in the $k$-th bucket is denoted as $y_{k} = \mathbf{1}[\Delta_t > \tau_k]$, and the model is optimized using a Binary Cross-Entropy (BCE) loss across all buckets:
$$ \mathcal{L}_{next} = \sum_{k=1}^{K}BCE(y_{k}, z_{k}). $$

% \textbf{Online Inference.} During online inference, we decode the ordinal logits into a precise continuous time gap by smoothly aggregating the probabilities:
% $$\Delta_t = \sum_{k=1}^{K}(\tau_k - \tau_{k-1}) \cdot \text{Sigmoid}(z_{k}).$$In addition, we can adjust the RTG to achieve different desired business values. We expect the predicted $\Delta_t$ to work well under different input RTG. Therefore, we need to make $\Delta_t=\pi_P(s_t,R_t)$ capture the effect of $\lambda_n$ in $R_t$. To achieve this, instead of fixing $\lambda_n$, we sample $\lambda_n$ from a uniform distribution $U(0,1)$ in the training stage. As a result, we can adjust the $\lambda_n$ to get the desired business effect when inferring online.
\noindent
\textbf{Online Inference.} During online inference, we decode the ordinal logits into a precise continuous time gap by smoothly aggregating the probabilities with $\tau_0=0$:
$$ \Delta_t = \sum_{k=1}^{K}(\tau_k - \tau_{k-1}) \cdot \text{Sigmoid}(z_{k}). $$
Furthermore, to allow dynamic adjustments of business objectives during online inference, the predicted $\Delta_t$ must generalize well across different input RTG values. Therefore, the policy $\Delta_t=\pi_P(s_t,R_t)$ needs to capture the specific effect of $\lambda_n$ embedded in $R_t$. To achieve this, instead of fixing $\lambda_n$ during the training stage, we condition the policy on $\lambda_n$ and sample $\lambda_n$ from a uniform distribution $U(0,1)$ based on a previous work~\cite{abels2019dynamic}. As a result, we can adjust $\lambda_n$ during online inference to achieve the desired trade-off between positive and negative business effects without retraining. 

% In addition, a higher positive RTG is set for inactive users, and a lower positive RTG is set for active users.

\subsection{Execution Agent}

% The execution agent is responsible for determining whether to send a push at the time generated by the planning agent. Formally, based on $s_t$, inspired by GAVE \cite{gao2025generative_gave}, we propose a Generative Value-Guided Decision architecture, with the same Gated-RTG and $\lambda_n$ sampling to ensure the online performance tuning ability.
The execution agent is responsible for determining whether to send a push at the time generated by the planning agent. Formally, based on $s_t$, and inspired by GAVE~\cite{gao2025generative_gave}, we propose a Generative Value-Guided Decision architecture, incorporating the same Gated-RTG and $\lambda_n$ sampling mechanisms to enable dynamic online performance tuning.

\noindent
\textbf{Generative Value-Guided Decision.} Specifically, we need to estimate an action-value function $Q(s_t,a_t|\lambda_n)$ to learn $R_{t+1}$ alongside an action generation policy $\pi(a_t|s_t,R_t)$. Unlike the original GAVE framework, which involves action sampling in a continuous space, we omit this step since our push scenario features a discrete binary action space. Furthermore, the original GAVE paper directly regresses $Q(s_t,a_t|\lambda_n)$ on the observed RTG $R_t$ as the loss function. However, because the actions in offline training data may be suboptimal, the observed RTG fails to accurately represent the true future reward under the given state $s_t$ \cite{yamagata2023qdt}. Consequently, directly regressing on the observed RTG yields suboptimal performance. Thus, we propose using the Bellman equation to learn $Q(s_t,a_t|\lambda_n)$. Taking the positive reward $r^+_t$ as an example, the corresponding $Q$-value represents the user's active probability. Intuitively, an inactive user will have $0$ reward all the time, and an active user will have $1$ reward at time $M$ if the user become active between $M$ and $M+1$. Therefore, we define the reward for an active user as below:
$$ r^+_t = \begin{cases} 0, & t < M, \\ 1, & t = M \text{ and user active between $M$ and $M+1$,} \\ 0, & t > M. \end{cases} $$
Following this, the positive RTG is iteratively formulated as:
$$ R^+_{t} = \begin{cases} 0, & \text{if the user is already active}\\ r_t^+ + \gamma \max_a Q^+(s_t,a|\lambda_n), & \text{via the Bellman equation} \end{cases} $$
This identical technique applies to defining the negative RTG, $R^-_t$. We then learn $Q(s_t, a_t|\lambda_n)$ using Mean Squared Error (MSE) loss:
$$ \mathcal{L}_Q = \sum_{t=1}^{T-1}(R_{t+1} - Q(s_t, a_t|\lambda_n))^2, $$
where $R_t = R_t^+ - \lambda_n R_t^-,~ Q(s_t, a_t|\lambda_n) = Q^+(s_t, a_t|\lambda_n) - \lambda_n Q^-(s_t, a_t|\lambda_n)$. Our goal is to guide the policy to learn better actions corresponding to higher $Q(s_t, a_t|\lambda_n)$ values. Following GAVE~\cite{gao2025generative_gave}, we calculate the advantage weight $w_t = \text{Sigmoid}(\alpha_r \cdot (Q(s_t, a_t|\lambda_n) - Q(s_t, 1-a_t|\lambda_n)))$ to reweight the action loss below, where $\alpha_r$ is a scaling hyperparameter: 
$$ \mathcal{L}_\pi = - \sum_{t=1}^{T}\left[ w'_t \log \pi(a_t|s_t,R_t) + (1-w'_t) \log(1- \pi(a_t|s_t,R_t)) \right], $$
where $w'_t$ represents the weight with gradients frozen. The intuition here is to assign higher weights to actions with larger relative $Q$-values, thereby explicitly guiding the execution agent toward optimal behaviors.

To summarize, the final loss function for model training is
$$\mathcal{L}=\mathcal{L}_{next}+\lambda_1\mathcal{L}_{Q}+\lambda_2\mathcal{L}_{\pi}.$$

% \textbf{Discussion of Model Structure.} In this subsection, we discuss the model structure difference between the planning and execution agent. First, since predicting the exact next delivery time is a high-noise task with a continuous action space, we use the traditional decision transformer structure based on imitation learning to ensure a lower bound of performance. Specifically, this high level of noise arises because the planning agent sits upstream in the push system. The effect of each single decision can be diluted by multiple factors. In addition, the push time decisions do not directly determine delivery, leading to a relatively weak connection between RTG and generated push time. Meanwhile, we find that introducing a critic can produce a higher performance upper bound empirically, but it is more difficult to tune parameters and may easily suffer from reward hacking. Therefore, we use this structure in the binary execution scenario with less noise.
\noindent
\textbf{Discussion of Model Structure.} In this subsection, we discuss the model structure difference between the planning and execution agents. First, since predicting the exact next delivery time is a high-noise task with a continuous action space, we use the traditional decision transformer structure to ensure strong system robustness. Specifically, this high level of noise arises because the planning agent sits upstream in the push system, and the effect of each single decision can be diluted by multiple factors. In addition, the push time decisions do not directly determine final delivery, leading to a relatively weak connection between the RTG and generated push time. Meanwhile, we find that introducing a critic can produce a higher performance upper bound empirically, but it is more difficult to tune parameters and may easily suffer from reward hacking under high-noise conditions. Therefore, instead of using it for the planning agent, we use this value-guided structure exclusively in the binary execution scenario with less noise.

\begin{figure}[t]
\centering
% \vspace{-5pt}
{
\begin{minipage}[t]{\linewidth}
\centering
\includegraphics[width=\textwidth]{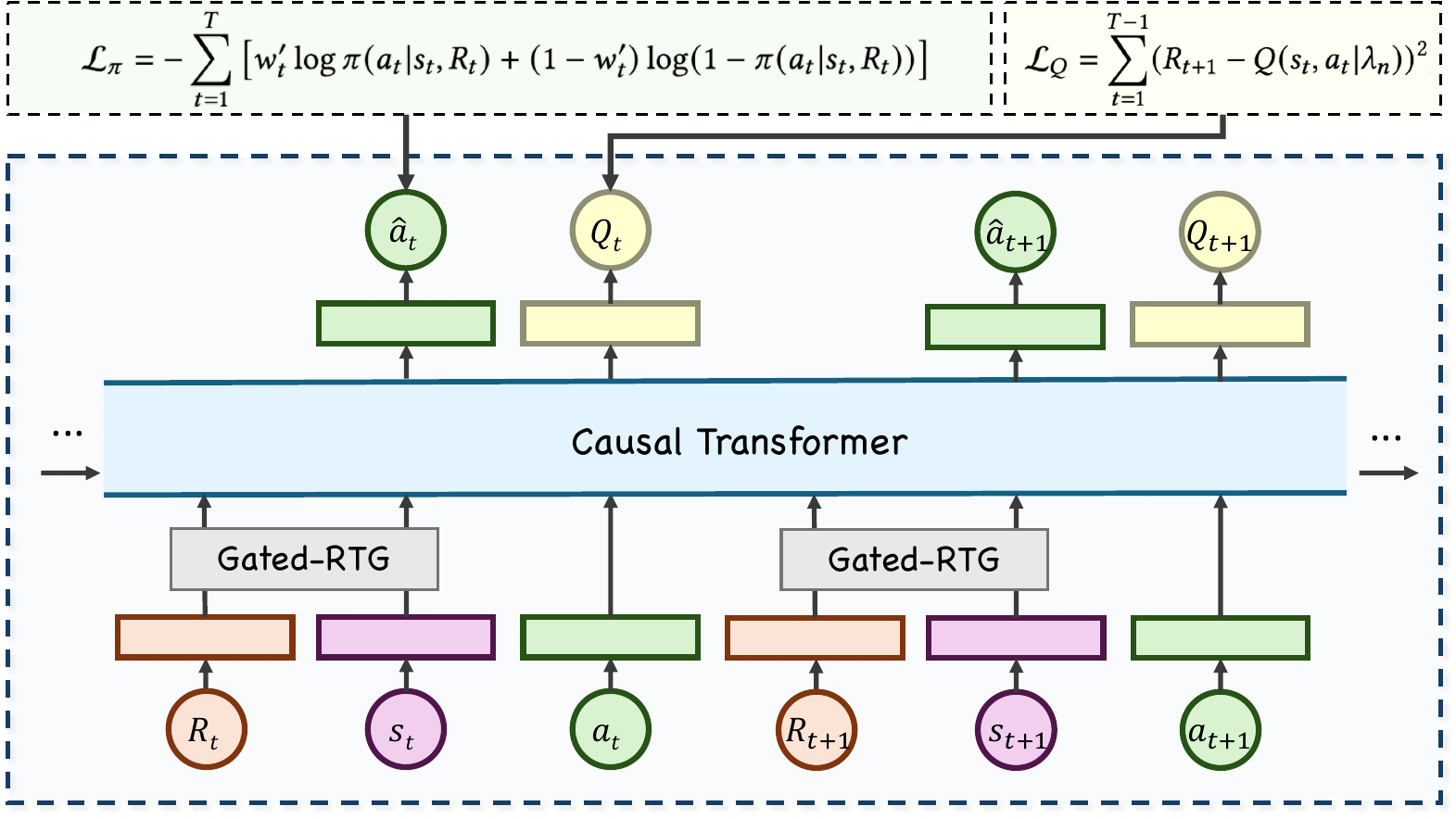}
\end{minipage}%
}%
\\
\centering
\caption{Architecture of the execution agent.}
% \vspace{-8pt}
\label{fig:Execution Agent}
\end{figure}

\subsection{System Landing Challenge}
% First, given the large volume of business events we have, such as live streams and video posts, relying solely on the planning agent carries the risk of missing critical events. For example, if a star followed by a user goes live, this is a potential push time. Therefore, in our implementation, we integrate such business events with self‑generated push time to better enhance user experience.
First, given the massive volume of real-time business events on our platform (e.g., followed creators live starts and new video posts from friends), relying exclusively on the planning agent risks missing critical, time-sensitive occurrences. For instance, if a creator followed by a user suddenly goes live, this represents a high-potential, immediate push opportunity. Therefore, in our practical implementation, we seamlessly integrate these event-driven triggers with the agent-generated timestamps to ensure a responsive and comprehensive user experience.

% Second, our platform handles millions of queries per second (QPS). Since our system does not rely on fixed budget limits and pre-defined delivery time periods, there is a high risk of unreasonable planning, such as too short or too long. In addition, if the sending is too frequent, it will consume too many computing resources. Therefore, we deploy a lightweight filtering agent to ensure the safety and efficiency of the entire agentic push system.
Second, our platform handles millions of queries per second (QPS). Because our system discards fixed budget limits and pre-defined delivery windows, there is an inherent risk of anomalous temporal planning (e.g., generating intervals that are excessively short). Furthermore, hyperactive system triggering would consume immense computing resources. To mitigate this, we introduce a lightweight filtering agent to ensure the safety, stability, and resource efficiency of the entire agentic pipeline.

% To avoid extreme user experiences, we introduce two primary boundary controls:
% \begin{itemize}
% \item System-side control: This rule suppresses the activation of the downstream system for push time within a fixed window after the most recent successful send.

% \item User-side control: This rule suppresses the activation for push time within a fixed window after the user becomes active.

% \end{itemize}

To prevent severe negative user experiences, we introduce two primary boundary controls:
\begin{itemize}
\item \textbf{System-side control:} Suppresses downstream activation for a fixed window immediately following a recently successful push delivery.
\item \textbf{User-side control:} Suppresses downstream activation for a fixed window immediately after a user naturally becomes active on the platform.
\end{itemize}

% In practice, these boundary controls are set to highly conservative, hardcoded values on the scale of minutes rather than hours simply to prevent extreme user experiences. Beyond these basic safeguards, the primary filtering decision relies on a lightweight model based solely on user and environment information. Precisely, we distill the knowledge from the heavy action execution agent into a 3-layer MLP without item-side features. This exclusion is crucial because incorporating item-side features would require invoking the computationally expensive downstream item sorting module. By distilling the model decisions and dropping these item-side features, we find a slight degradation in model metrics, but a massive improvement in resource efficiency.
In practice, these boundary controls use highly conservative, hardcoded thresholds (on the scale of minutes rather than hours) strictly to prevent edge-case notification spam. Beyond these basic safeguards, the core filtering decision is governed by a lightweight model relying entirely on user and environment features. Specifically, we distill the knowledge from the heavy execution agent into a 3-layer MLP, intentionally omitting item-side features. This exclusion is critical: incorporating item features would inevitably invoke the downstream item-sorting pipeline. Because the complex stages of item recall, pre-ranking, and ranking dominate the computational overhead, this full pipeline consumes roughly $10\times$ more resources compared to using user and environment features alone. By distilling the decision process and acting as a robust gatekeeper, we accept a marginal degradation in model metrics in exchange for preventing the system from wasting this massive compute on low-potential push timings.

% \textbf{Computational Resource Efficiency.} In industrial push recommendation pipelines, the downstream execution stage is computationally heavy, since it requires complex item recall, pre-ranking, and ranking using massive item-side features. In our scenario, including item sorting consumes $10$ times more resources than only using user features and environment. The proposed filtering agent can also prevent the system from wasting valuable compute on low-potential push timing. We will discuss the specific resource savings in the experimental section.

\section{Experiment}
% We train the model on the Douyin production datasets and conduct experiments on online traffic. The offline training and evaluation are conducted on a dataset comprising over six months of logs from more than one billion users, ensuring the model's robustness and generalizability across diverse user segments. All online experiments are conducted as A/B tests on the Douyin push notification platform, ensuring completely random assignment of user devices, and have lasted for 14 days.
We evaluate our proposed method using both offline industrial datasets and online A/B testing. The offline training and evaluation utilize a massive dataset comprising over six months of production logs from more than one billion Douyin users, ensuring the model's robustness and generalizability across diverse user segments. The online A/B tests are conducted directly on the Douyin push notification platform with completely randomized user device assignments, running for 14 consecutive days.

\subsection{Baseline and Experiment Protocol}
To demonstrate the effectiveness of our proposed method, we compare it against two representative state-of-the-art paradigms:
% \begin{itemize}
%     \item \textbf{Pre-Planned Frequency:} Our online production push system based on the pre-planned frequency strategy. Before the day begins, this multi-stage system relies on uplift modeling to pre-plan the push frequency and allocates delivery timings for each user in advance via integer programming solvers, operating under a global budget constraint~\cite{li2026user}.
%     \item \textbf{Fixed-Interval Triggering~\cite{yuan2025generative_linkedin}:} An industrial Decision Transformer-based model that bypasses the timing issue by implementing a fixed-interval triggering mechanism to determine \textit{whether} to send a notification. To ensure a fair online A/B comparison, we carefully adjusted its polling frequency so that its resource consumption is practically identical to the Baseline.
% \end{itemize}
\begin{itemize}
    \item \textbf{Pre-Planned Frequency~\cite{li2026user}:} The current production baseline at Douyin. Before the target day begins, this multi-stage system relies on uplift modeling to pre-plan the push frequency and allocates delivery timings for each user in advance via integer programming solvers under a global budget constraint.
    \item \textbf{Fixed-Interval Triggering~\cite{yuan2025generative_linkedin}:} An industrial Decision Transformer-based model that bypasses exact timing generation by implementing a fixed-interval mechanism to periodically determine \textit{whether} to send a notification. To ensure a fair online A/B comparison, we carefully tuned its polling frequency to align its resource consumption as closely as possible with the baseline.
\end{itemize}
% We report relative percentage changes compared to the production baseline. We measure the business value across three critical dimensions: \textbf{User Active Days (UAD)} (the higher the better), \textbf{Negative Experience (NE)}, and \textbf{Resource Consumption} (computational cost, the lower the better). Specifically, we use the number of users who disable push notifications during the A/B test as a metric for NE.
We report the relative percentage changes compared to the pre-planned frequency baseline across three critical dimensions: \textbf{User Active Days (UAD)} ($\uparrow$), \textbf{Negative Experience (NE)} ($\downarrow$), and \textbf{Resource Consumption} ($\downarrow$). Specifically, NE is quantified by the number of users who revoke their push notification permissions during the A/B test; this metric serves as a direct indicator of user fatigue caused by excessive or poorly timed interruptions. Resource Consumption reflects the overall computational cost of the system.

% 对比SOTA
\begin{table}[tbp]
\centering
\caption{Comparison with state-of-the-art methods. All metric values denote the relative percentage changes compared to the baseline.}
\label{tab:sota}
\begin{tabular}{lccc}
\toprule
\textbf{Method} & \textbf{UAD $\uparrow$} & \textbf{NE $\downarrow$} & \textbf{Resource $\downarrow$} \\
\midrule
Pre-planned frequency   & - & - & - \\
Fixed-interval triggering  & -0.0670\% & +0.0205\% & +6.548\% \\
\textbf{STEPS}               & \textbf{+0.2843\%} & \textbf{-1.9089\%} & \textbf{-79.42\%} \\
\bottomrule
\end{tabular}
\end{table}

% As shown in Table~\ref{tab:sota}, our self-triggered agentic system significantly outperforms both paradigms. The fixed-interval approach struggles to surpass the pre-planned baseline under the matched resource budget, resulting in a slight degradation in UAD (-0.0670\%) and an increase in NE (+0.0205\%). In contrast, our approach achieves a +0.2843\% increase in UAD and a -1.9089\% reduction in NE. More importantly, by reformulating the problem into a closed-loop framework where the system dynamically predicts its own next activation time, we eliminate the waste associated with fixed time slices and the rigidity of offline frequency planning. It is worth noting that while the self-triggered mechanism intrinsically avoids redundant polling, this massive 79.42\% resource reduction is primarily realized by our newly introduced lightweight filtering agent (as further detailed in Section 4.2), which efficiently prunes low-value requests before they reach the compute-intensive execution stage.
As shown in Table~\ref{tab:sota}, our STEPS framework significantly outperforms both existing paradigms. Under a comparable resource budget, the fixed-interval approach struggles to surpass the pre-planned baseline, resulting in a slight degradation in UAD (-0.0670\%) and an increase in NE (+0.0205\%). In contrast, our approach achieves a +0.2843\% increase in UAD and a substantial -1.9089\% reduction in NE. More importantly, by reformulating the problem into a closed-loop framework where the system dynamically predicts its own next activation time, we eliminate the rigidity of offline frequency planning and the computational waste of fixed time slices. It is worth noting that while this self-triggered mechanism intrinsically avoids redundant polling, the massive 79.42\% resource reduction is primarily realized by our newly introduced lightweight filtering agent (detailed in Section \ref{sec:Ablation}), which efficiently prunes low-value requests before they invoke the compute-intensive downstream item-sorting and execution pipeline.

\subsection{Ablation Study}\label{sec:Ablation}
% 总体表格，统计了三个大模块各自的贡献
\begin{table}[tbp]
\centering
\caption{Ablation study on the contribution of each agent.}
\label{tab:ablation}
\begin{tabular}{lccc}
\toprule
\textbf{Agent} & \textbf{UAD $\uparrow$} & \textbf{NE $\downarrow$} & \textbf{Resource $\downarrow$} \\
\midrule
Planning         & +0.1808\% & -0.9781\% & -4.54\% \\
Execution & +0.1035\% & -0.6843\% & - \\
Filtering        & - & -0.2465\% & -74.88\% \\
\bottomrule
\end{tabular}
\end{table}

% To quantify the overall and respective contribution of the three core components (\textbf{planning}, \textbf{execution}, and \textbf{Filtering}), we report results in Table~\ref{tab:ablation}. The results indicate that all three agents effectively and synergistically drive the system's performance, collectively achieving a 0.2843\% increase in UAD, 1.9089\% reduction in Negative Experience (NE), and 79.42\% reduction in computational resource. Among the three agents, the planning agent serves as the dominant contributor to both the positive and negative metrics. It accounts for the largest share of the overall positive gain (+0.1808\% UAD) and over half of the total negative-metric reduction (-0.9781\% NE). This demonstrates that better planning not only ensures valuable opportunities are properly captured but also proactively avoids unreasonable sending times that cause user disturbance.
To quantify the respective contributions of the three core components (planning, execution, and filtering), we report results in Table~\ref{tab:ablation}. The results indicate that all three agents effectively and synergistically drive the system's performance. Among them, the planning agent serves as the dominant contributor to both the positive and negative metrics (+0.1808\% UAD, -0.9781\% NE). This demonstrates that better planning not only ensures valuable opportunities are properly captured but also proactively avoids unreasonable sending times that cause user disturbance. Complementing this, the execution agent further amplifies these gains (+0.1035\% UAD, -0.6843\% NE) by accurately evaluating candidate items at the planned times. Finally, the filtering agent provides boundary controls to further suppress NE (-0.2465\%), while its pruning of low-potential requests accounts for a massive -74.88\% reduction in computational resources.

% \begin{figure}[t]
% \centering
% % \cente
% \includegraphics[width=0.9\linewidth]{figures/next_trigger_image1.png}
% \\
% \centering
% \caption{Effectiveness of learning correlation between RTG and action.}
% \label{fig:corr_rtg_act}
% \end{figure}

\begin{figure}[t]
    \centering

    \begin{subfigure}[t]{0.23\textwidth}
        \centering
        \includegraphics[width=\linewidth]{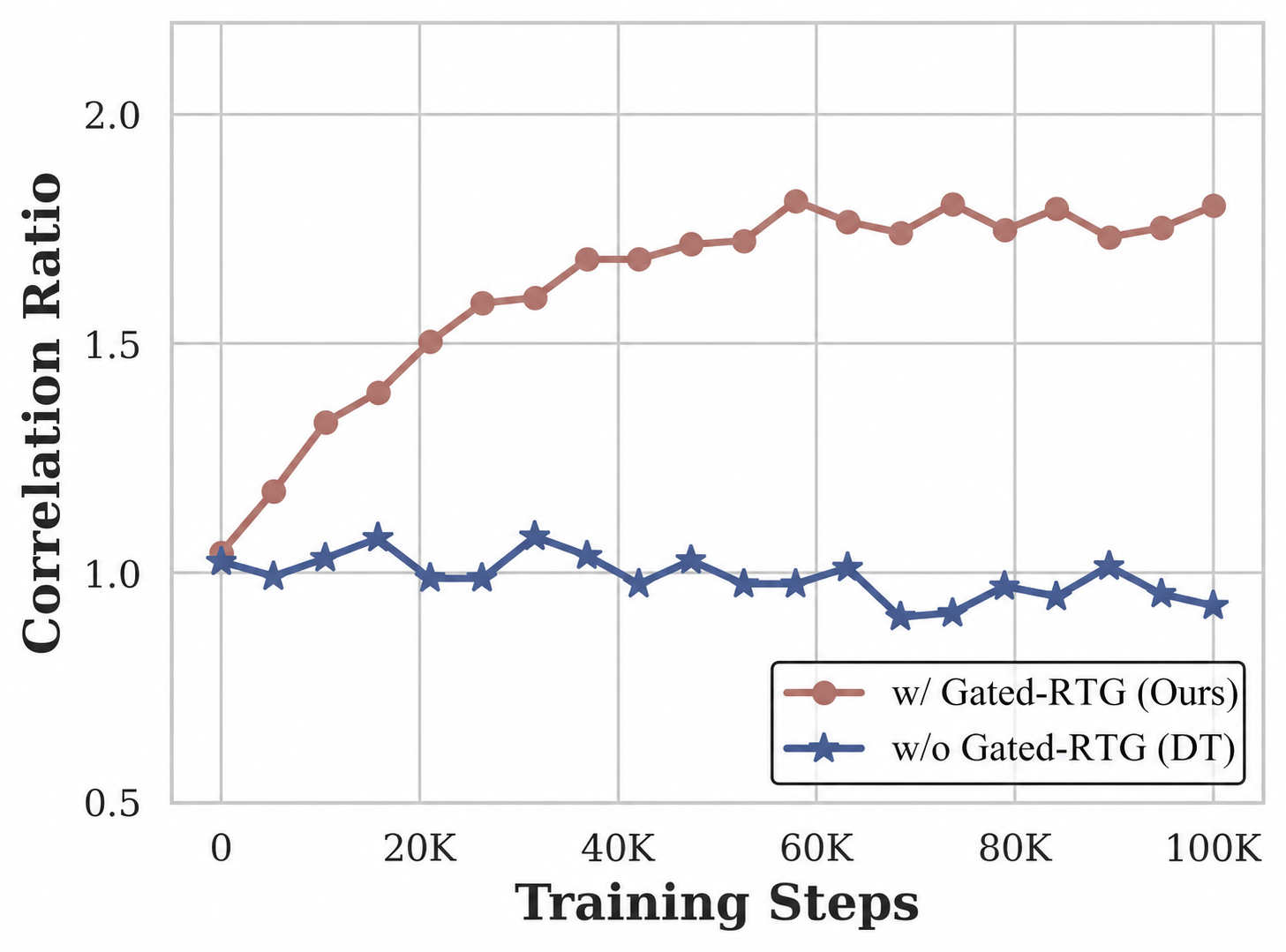}
        \caption{Effectiveness of learning correlation between RTG and action.}
        \label{fig:corr_rtg_act}
    \end{subfigure}
    \hfill
    \begin{subfigure}[t]{0.23\textwidth}
        \centering
        \Description{Offline regression AUC under equal-frequency and equal-width ordinal bucket configurations. Equal-frequency buckets perform best at one hundred buckets, which is the deployed configuration.}
        \includegraphics[width=\linewidth]{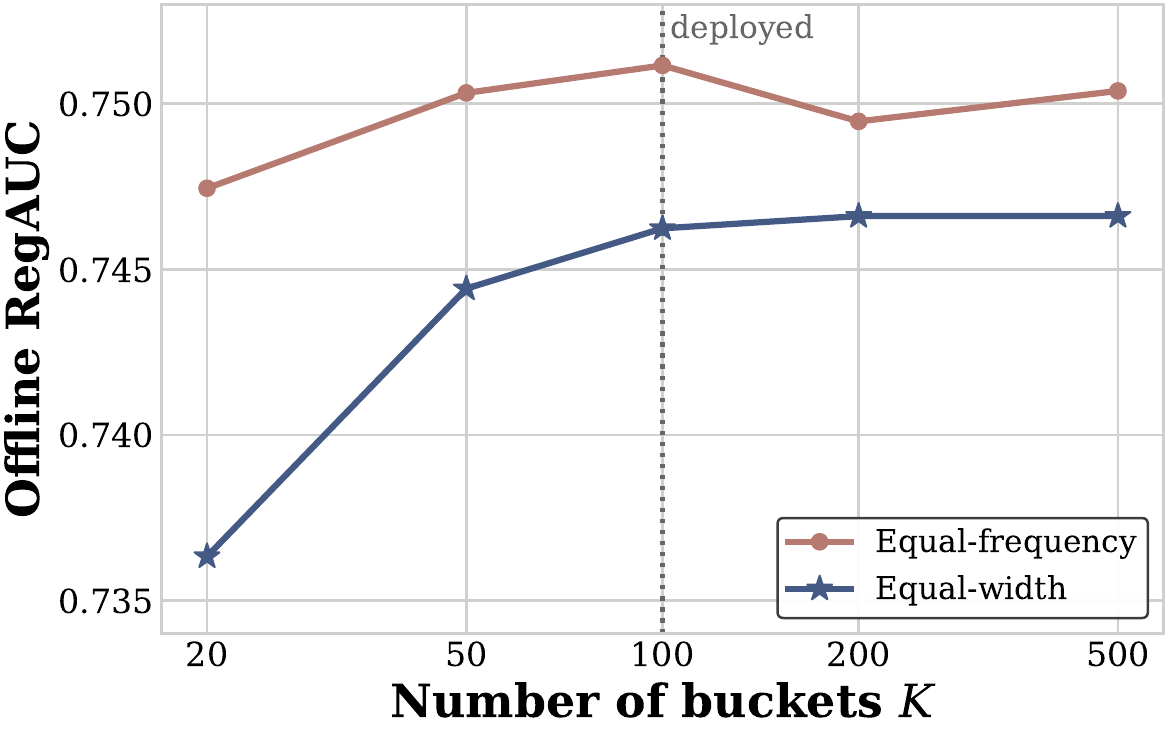}
        \caption{Sensitivity to the ordinal bucket count and boundary strategy.}
        \label{fig:ordinal_bucket_sensitivity}
    \end{subfigure}

    \caption{Comparison of the effectiveness of RTG-action correlation learning and the sensitivity to bucket configuration.}
    \label{fig:combined_subfigures}
\end{figure}
\subsection{Further Analysis on Planning}

\noindent\textbf{Gated RTG Structure.} In our highly noisy industrial environment, the observed RTG exhibits extreme variance, often causing standard conditional models to ignore this signal altogether. To quantify whether the model actually follows the RTG prompt when predicting the next activation time gap $\Delta_t$, we define the \textit{Correlation Ratio}:
$$
\text{Correlation Ratio} \;=\; \frac{\mathbb{E}_{x}\!\left[\hat{\Delta}_t(\text{RTG}=r_{\text{low}})\right]}
              {\mathbb{E}_{x}\!\left[\hat{\Delta}_t(\text{RTG}=r_{\text{high}})\right]}
$$
where $r_{\text{low}}=0.0$ and $r_{\text{high}}=1.0$. A ratio near $1.0$ indicates a severe ``condition-ignoring'' problem—meaning the model outputs the same action regardless of the target RTG—whereas a ratio clearly deviating from $1.0$ proves the model has established a stable condition-action alignment.

As illustrated in Figure~\ref{fig:corr_rtg_act}, the standard Decision Transformer without Gated-RTG fails this evaluation, with its correlation ratio randomly fluctuating around 1.0. Because standard concatenation treats the 1D, high-variance RTG scalar as just another input feature, the neural network optimizes by simply bypassing it, allowing the fragile condition signal to be completely overwhelmed by the rich, high-dimensional state embeddings. In contrast, our proposed Gated-RTG mechanism multiplicatively forces the state embeddings to be modulated by the RTG. By eliminating the network's ability to bypass this unreliable condition, the ratio steadily converges to approximately 1.8—a value closely matching the true empirical statistics of our training data. This strongly demonstrates that the Gated-RTG design effectively overcomes feature-overwhelming, robustly capturing the underlying data distribution despite severe environmental noise.

\begin{table}[tbp]
\centering
\caption{Relative change of pass rate after applying the planning agent.}
\resizebox{0.8\linewidth}{!}{
\begin{tabular}{cc}
\toprule
\textbf{Filtering Pass Rate} & \textbf{Execution Pass Rate} \\
\midrule
+12.3\% & +2.3\% \\
\bottomrule
\end{tabular}
}
\label{tab:downstream_passrate}
\end{table}

% \noindent\textbf{Downstream Pass Rate.} Ideally, an optimal generated push time should successfully pass through all downstream agents to trigger a delivery, thereby maximizing both business value and computational resource utilization. As reported in Table~\ref{tab:downstream_passrate}, the pass rate of Filtering and Execution improved significantly. This is mainly because our RTG is collected end-to-end from the planning agent's decision through Filtering, Execution, and the final user response. Consequently, any block contributes zero to the RTG, making the Filtering outcome an implicit supervisory signal embedded in the reward. Training the planning agent on this RTG therefore biases its outputs toward time gaps that survive downstream filtering. The same compute resources now sustain a higher volume of high-quality requests. We identify this as the
% primary driver behind the observed gain in user active days.
\noindent\textbf{Downstream Pass Rate.} An optimal generated push time should seamlessly pass through all downstream agents to trigger a delivery, maximizing both business value and resource efficiency. As reported in Table~\ref{tab:downstream_passrate}, the pass rates for both the filtering and execution agents improve significantly. This is primarily because our RTG is collected end-to-end—spanning from the initial planning decision, through the downstream agents, to the final user response. Consequently, any request blocked by downstream agents yields a zero RTG. This embeds the filtering outcome into the reward as an implicit supervisory signal. By optimizing for this end-to-end RTG, the planning agent is naturally biased toward generating time gaps that survive downstream evaluation. As a result, the same computational budget now sustains a higher volume of high-quality, deliverable requests, which we identify as the primary driver behind the observed gain in UAD.

\begin{table}[tbp]
\centering
\caption{Relative change of inter-push gap.}
\begin{tabular}{lcccc}
\toprule
Inter-push Gap &\textbf{0\textasciitilde20min} & \textbf{20min\textasciitilde1h} & \textbf{1h\textasciitilde3h} & \textbf{3h\textasciitilde6h} \\
\midrule
Change&-35.93\% & +20.37\% & +91.93\% & +179.84\% \\
\midrule
\midrule
Quantile & \textbf{0.01} & \textbf{0.05} & \textbf{0.25} & \textbf{0.5} \\ \midrule
Change & +0.1\%& +13.1\%& +20.1\%& +25.6\% \\
\bottomrule
\end{tabular}
\label{tab:next_trigger}
\end{table}
% \noindent\textbf{Sending Density.}
% We investigate whether the planning agent improves the distribution of sends, i.e., whether users perceive a more reasonable delivery pattern. Table~\ref{tab:next_trigger} reports the inter-push gap distribution.
% The planning agent substantially suppresses the dense-push tail---the
% fraction of short-interval pushes drops sharply, while that of
% longer-interval pushes rises---and consequently shifts the 5th--50th
% percentiles of the gap distribution toward larger values. By reducing dense pushes, the planning agent contributes to mitigating user negative experience.

%\textbf{Sending Density.} We investigate whether the planning agent improves the distribution of sends, i.e., whether users perceive a more reasonable delivery pattern. The results are shown in Table~\ref{tab:next_trigger}. The sending frequency for short intervals drops substantially, while that for longer intervals increases markedly. This indicates that the planning agent substantially suppresses the high-frequency tail that is the primary source of negative experience. For the sending quantile, since the extreme frequent pushes is largely reduced, therefore the push time for $5$-th to $50$-th percentiles increases compared to the baseline.

\noindent\textbf{Sending Density.} We investigate whether the planning agent establishes a more reasonable delivery pattern from the user's perspective. Table~\ref{tab:next_trigger} reports the changes in the inter-push gap distribution. The results demonstrate that the planning agent substantially suppresses high-frequency bursts. Specifically, the fraction of extreme short-interval pushes (0--20 min) drops sharply by 35.93\%, while the proportion of longer, well-spaced intervals (3--6 h) rises massively by 179.84\%. Consequently, the 5th to 50th percentiles of the gap distribution consistently shift toward larger values. By proactively curtailing dense push clusters, the planning agent acts as the key mechanism for mitigating user fatigue and reducing NE.
% \noindent

% \begin{figure}[t]
%     \centering
%     \Description{Offline regression AUC under equal-frequency and equal-width ordinal bucket configurations. Equal-frequency buckets perform best at one hundred buckets, which is the deployed configuration.}
%     \includegraphics[width=0.9\linewidth]{figures/ordinal_bucket_sensitivity.pdf}
%     \caption{Sensitivity to the ordinal bucket count and boundary strategy. The vertical dotted line denotes the deployed $K=100$ configuration.}
%     \label{fig:ordinal_bucket_sensitivity}
% \end{figure}

% \noindent\textbf{Ordinal Regression Bucket Configuration.} We use ordinal regression to ensure more stable prediction of the continuous time gap $\Delta_t$. We compare two types of bucket configuration, i.e., (a) Equal-width: each bucket has the same width, and (b) Equal-frequency: each bucket covers the same number of samples. For each type, we choose difference value of $K$ from $\{20,50,100,200,500\}$.  Figure~\ref{fig:ordinal_bucket_sensitivity} shows that equal-frequency consistently outperform equal-width because they allocate resolution to the dense, short-gap region. With equal-frequency buckets, performance is stable from $K=50$ to $500$, but peaks at $K=100$ (with Regression AUC 0.75116). A smaller $K$ loses temporal resolution, while a larger $K$ sparsifies supervision without bringing further gains. We therefore use $K=100$ equal-frequency buckets in our system.
\noindent\textbf{Hyperparameter Analysis: Bucket Configuration.} To optimize the ordinal regression for predicting the continuous time gap $\Delta_t$, we conduct a hyperparameter analysis on the discretization strategy. We evaluate two bucket types (equal-width vs. equal-frequency) across varying bucket counts $K \in \{20, 50, 100, 200, 500\}$. As shown in Figure~\ref{fig:ordinal_bucket_sensitivity}, the equal-frequency strategy consistently outperforms equal-width by allocating finer resolution to dense, short-gap regions. Regarding the sensitivity to $K$, performance remains stable for $K \in [50, 500]$ under the equal-frequency setup, peaking at $K=100$ (Regression AUC 0.75116). Since a smaller $K$ degrades critical temporal resolution and a larger $K$ over-sparsifies the supervision signal without yielding further gains, we finalize our system configuration with the optimal hyperparameters of $K=100$ equal-frequency buckets.

% \begin{table}[t]
% \centering
% \caption{Sensitivity to the target RTG. ``Protected'' denotes the fraction of users with the highest close-notification propensity whose target RTG is reset to zero. All metrics are relative changes against their respective controls.}
% \label{tab:rtg_target_sensitivity}
% % \resizebox{\linewidth}{!}{
% \begin{tabular}{lcccc}
% \toprule
% \textbf{Target} & \textbf{Protected} & \textbf{UAD $\uparrow$} & \textbf{NE $\downarrow$} & \textbf{Send} \\
% \midrule
% High   & 0\%  & +0.0119\% & +1.0419\% & +0.2051\% \\
% Medium & 10\% & +0.0067\% & +0.8978\% & +0.3694\% \\
% Low    & 20\% & +0.0113\% & +0.4555\% & +0.2518\% \\
% \bottomrule
% \end{tabular}
% \end{table}

% \noindent\textbf{Sensitivity of the target RTG level.} Table~\ref{tab:rtg_target_sensitivity} evaluates the online performance of three different target RTG levels. Re-assigning a zero target to the top 20\% close-sensitive users reduces the NE increase from 1.0419\% to 0.4555\%, while retaining nearly the same LT gain as the unprotected high target (0.0113\% versus 0.0119\%). This demonstrates that the target RTG is a controllable safety knob: a lower target substantially improves the negative-experience trade-off without collapsing long-term value.

\subsection{Further Analysis on Execution}

% To explore the sources of benefits from the \textbf{execution agent}, we conducted further analysis.

% \begin{table}[t]
% \centering
% \caption{Pacer pass rate before and after adjustment, grouped by user activity status.}
% \begin{tabular}{c|c}
% \toprule
% Activity Status & Pacer Pass Rate \\
% \midrule
% Inactive & 83.9\% $\rightarrow$ 85.0\% \\
% Active   & 56.0\% $\rightarrow$ 50.4\% \\
% \bottomrule
% \end{tabular}
% \label{tab:pacer-pass-rate}
% \end{table}

% \begin{figure}[t]
% \centering
% % \cente
% \includegraphics[width=0.9\linewidth]{figures/time.jpeg}
% \\
% \centering
% \caption{Hourly execution agent pass rate.}
% \label{fig:time}
% \end{figure}

% \begin{figure}[t]
% \centering
% \centering
% \includegraphics[width=0.9\linewidth]{figures/trajectory.jpeg}
% \\
% \centering
% \caption{Comparison of positive value increase between request-level and trajectory-level reward.}
% \label{fig:trajectory-level}
% \end{figure}

\begin{figure}[t]
    \centering
    % 子图1：Hourly Pass Rate
    \begin{subfigure}{0.23\textwidth}
        \centering
        \includegraphics[width=\linewidth]{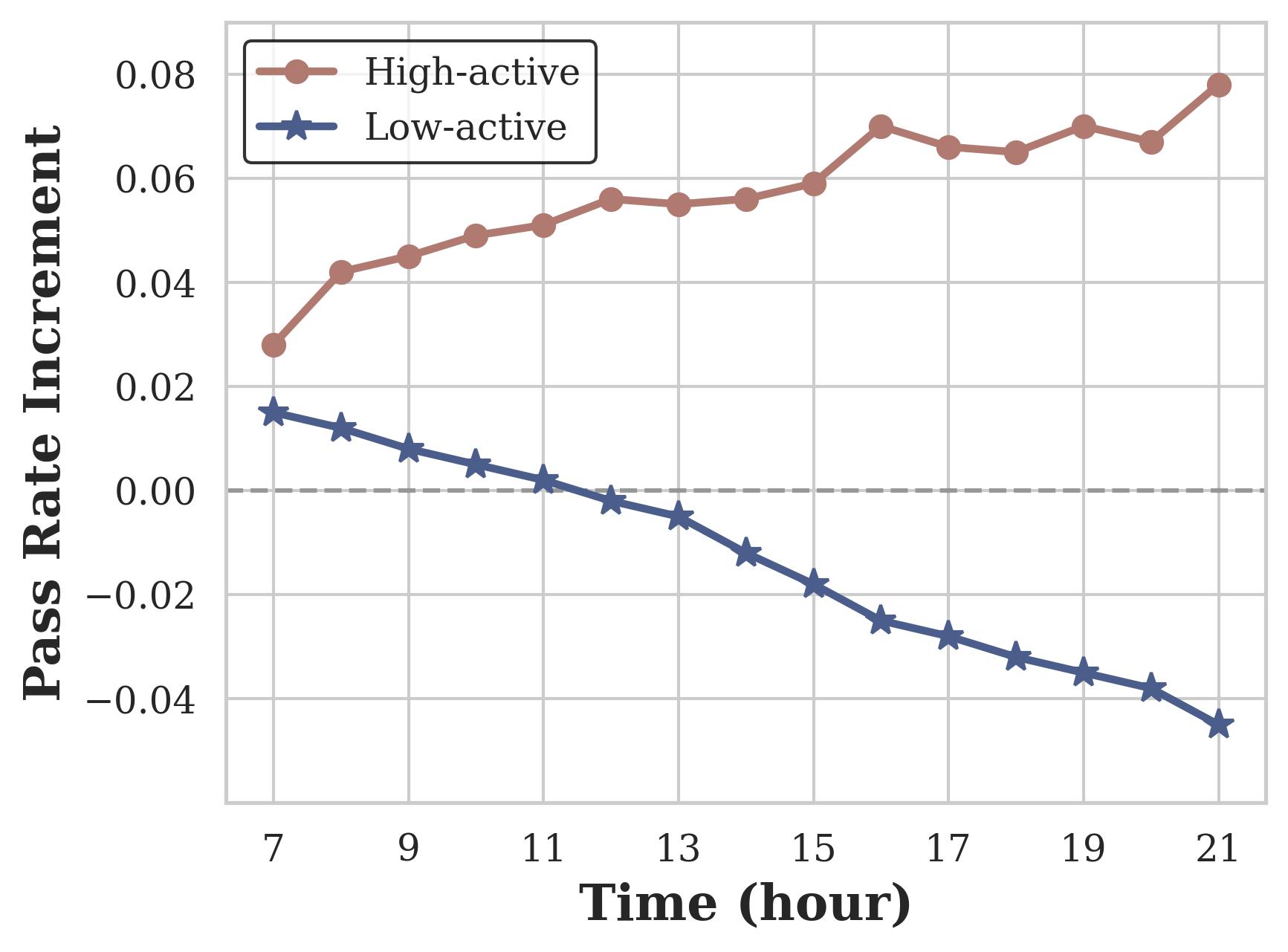}
        \caption{Hourly execution agent pass rate for currently inactive users.}
        \label{fig:time}
    \end{subfigure}\hfill
    % 子图2：Trajectory-level Reward
    \begin{subfigure}{0.23\textwidth}
        \centering
        \includegraphics[width=\linewidth]{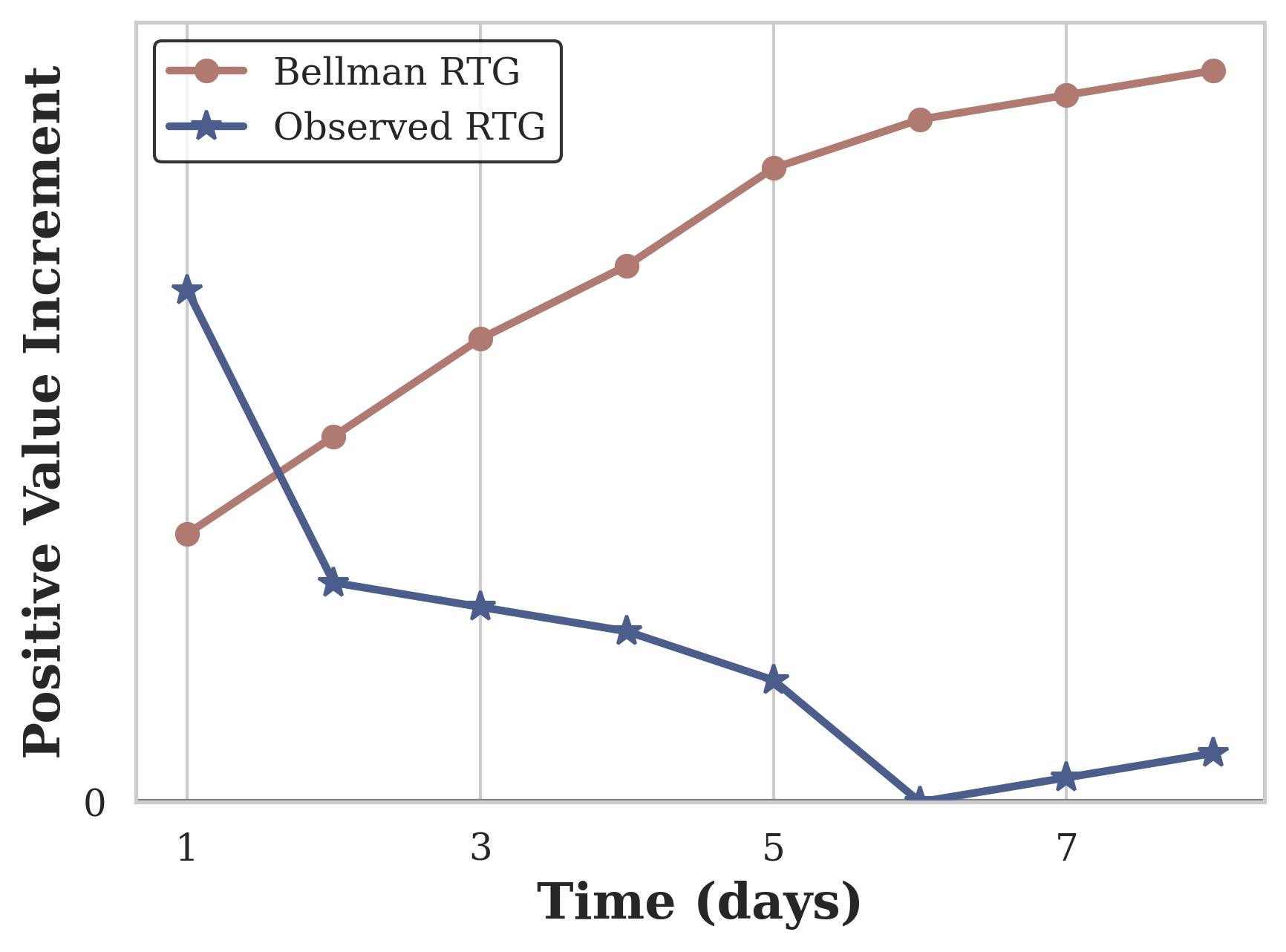}
        \caption{Comparison of positive value (UAD) increase between Bellman RTG and observed RTG.}
        \label{fig:trajectory-level}
    \end{subfigure}
    
    % 整个大图的总标题
    \caption{Analysis across push time and RTG calculation.}
    \label{fig:agent_performance}
\end{figure}

% \noindent
% \textbf{execution Pass Rate}

% Table~\ref{tab:pacer-pass-rate} reports the pass rate of the execution agent in the experimental group against the baseline, broken down by user activity state. The pass rate is lifted on inactive-state requests and simultaneously reduced on active-state requests, indicating that the execution agent reallocates the sending budget away from active-state requests and toward inactive-state ones. Since inactive-state requests carry substantially higher marginal activity days value, this asymmetric reallocation makes the overall decision policy more activity days-oriented.

\textbf{Intraday Pass Rate Dynamics.} To further demonstrate that the execution agent captures real-time dynamics, we analyze the increment of the pass rate across different hours. As shown in Figure~\ref{fig:time}, for currently inactive but historically high-engagement users, the pass rate increment steadily increases and remains positive, peaking in the evening. This indicates that if a high-active user remains inactive by night, sending a push yields substantial positive value increase. Conversely, for  currently inactive low-active users, the pass rate increment drops below zero into negative territory in the evening. This stark contrast reflects the model's trajectory awareness: if a low-active user remains unengaged by night, they are highly unlikely to become active, meaning the positive value of a push is minimal. These temporal phenomena support the rationality of the agent's execution strategy.

% \noindent\textbf{Long-term Value of Bellman RTG Modeling.} To validate whether the execution agent effectively captures the cumulative effects of push notifications, we monitor the long-term positive value increment on online A/B testing. As illustrated in Figure~\ref{fig:trajectory-level}, the traditional methods that directly learn $Q(s,a)$ on observed RTG experience an initial spike but rapidly degrade, with its positive value increment converging around zero over time. One possible reason is that the observed RTG may be suboptimal due to the wrong action in training data. In contrast, our proposed Bellman RTG modeling exhibits a continuous and steady rise in positive value increment, stabilizing consistently above the baseline. Note that both methods apply a Gated-RTG mechanism, showing the improvement can be attributed to the Bellman RTG method.
\noindent\textbf{Long-term Value of Bellman RTG Modeling.} To validate whether the execution agent effectively captures the cumulative effects of push notifications, we monitor the long-term positive value increment in online A/B testing. As illustrated in Figure~\ref{fig:trajectory-level}, traditional methods that directly learn $Q(s,a)$ from the observed RTG experience an initial spike but subsequently decline, with their positive value increment converging to near zero over time. This decline likely occurs because the observed RTG is inherently biased by suboptimal historical actions in the offline training data. In contrast, our proposed Bellman RTG modeling exhibits a continuous and steady rise, stabilizing consistently above the baseline. Since both evaluated methods employ the Gated-RTG mechanism, this sustained long-term improvement can be primarily attributed to the Bellman RTG formulation, demonstrating its effectiveness in optimizing cumulative trajectory value.

\begin{table}[t]
\centering
\caption{User-level sending distribution.}
\label{tab:send_distribution}
\begin{tabular}{lc}
\toprule
\textbf{Metric} & \textbf{Relative Change} \\
\midrule
Zero sends (0) & $-$4.06\% \\
Moderate sends (1$\sim$20) & $+$20.31\% \\
Frequent sends (21$\sim$30) & $-$6.87\% \\
Extreme sends ($>$30) & $-$9.30\% \\
\bottomrule
\end{tabular}
\label{tab:send_distribution}
\end{table}

% \begin{figure}[t]
%     \centering
%     % 第一张图：Calibration Plot
%     \begin{minipage}{0.23\textwidth}
%         \centering
%         \includegraphics[width=\linewidth]{figures/Q1.jpeg}
%         \caption{Calibration ratio of predicted $Q$ values.}
%         \label{fig:q_calibration}
%     \end{minipage}\hfill
%     % 第二张图：Q Diff Plot
%     \begin{minipage}{0.23\textwidth}
%         \centering
%         \includegraphics[width=\linewidth]{figures/Q2.jpeg}
%         \caption{Action value difference for unactivated users.}
%         \label{fig:q_diff}
%     \end{minipage}
% \end{figure}
\begin{figure}[t]
    \centering
    % 子图1：Calibration Plot
    \begin{subfigure}{0.23\textwidth}
        \centering
        \includegraphics[width=\linewidth]{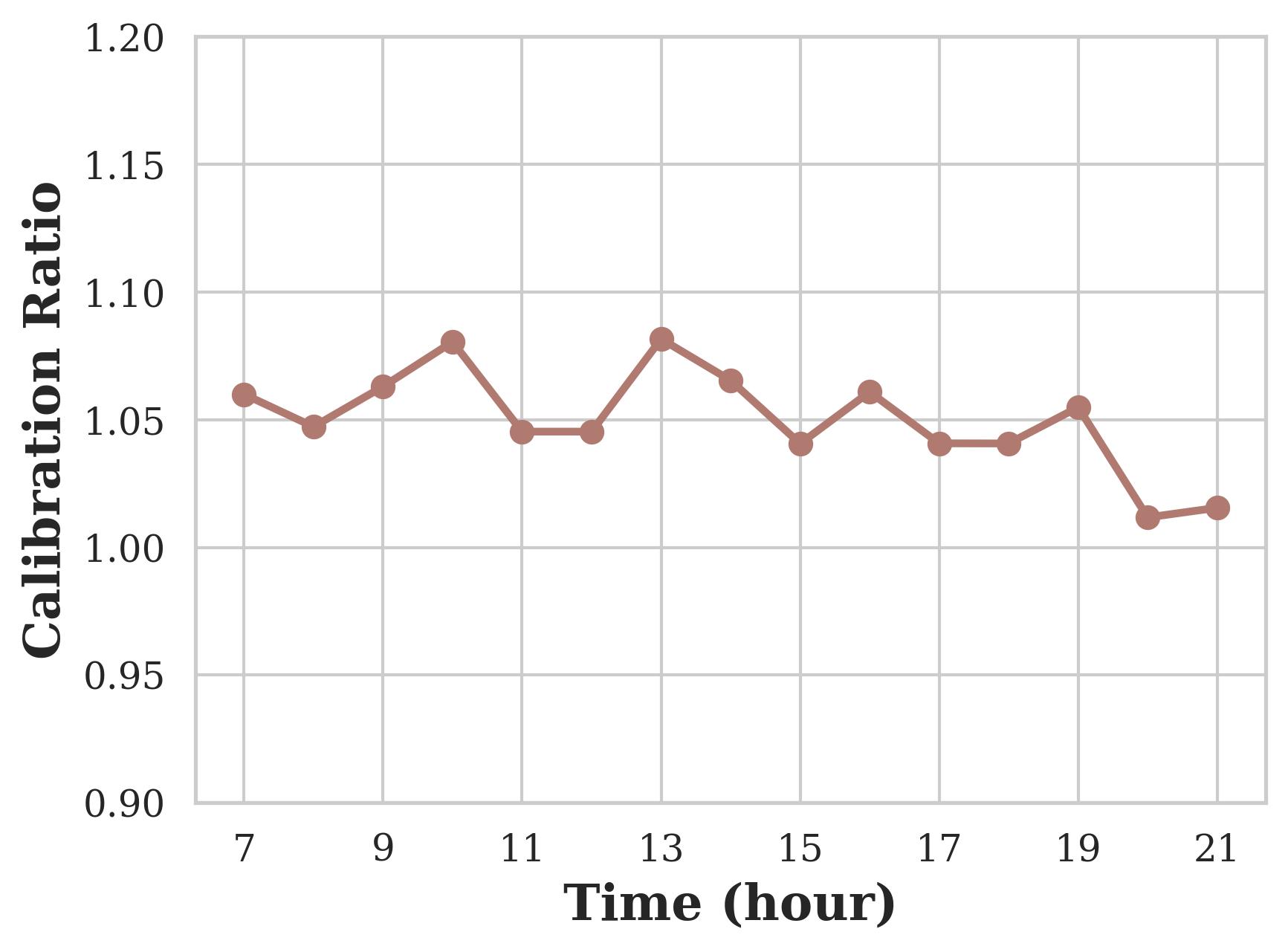}
        \caption{Calibration ratio of predicted $Q$ values.}
        \label{fig:q_calibration}
    \end{subfigure}\hfill
    % 子图2：Q Diff Plot
    \begin{subfigure}{0.23\textwidth}
        \centering
        \includegraphics[width=\linewidth]{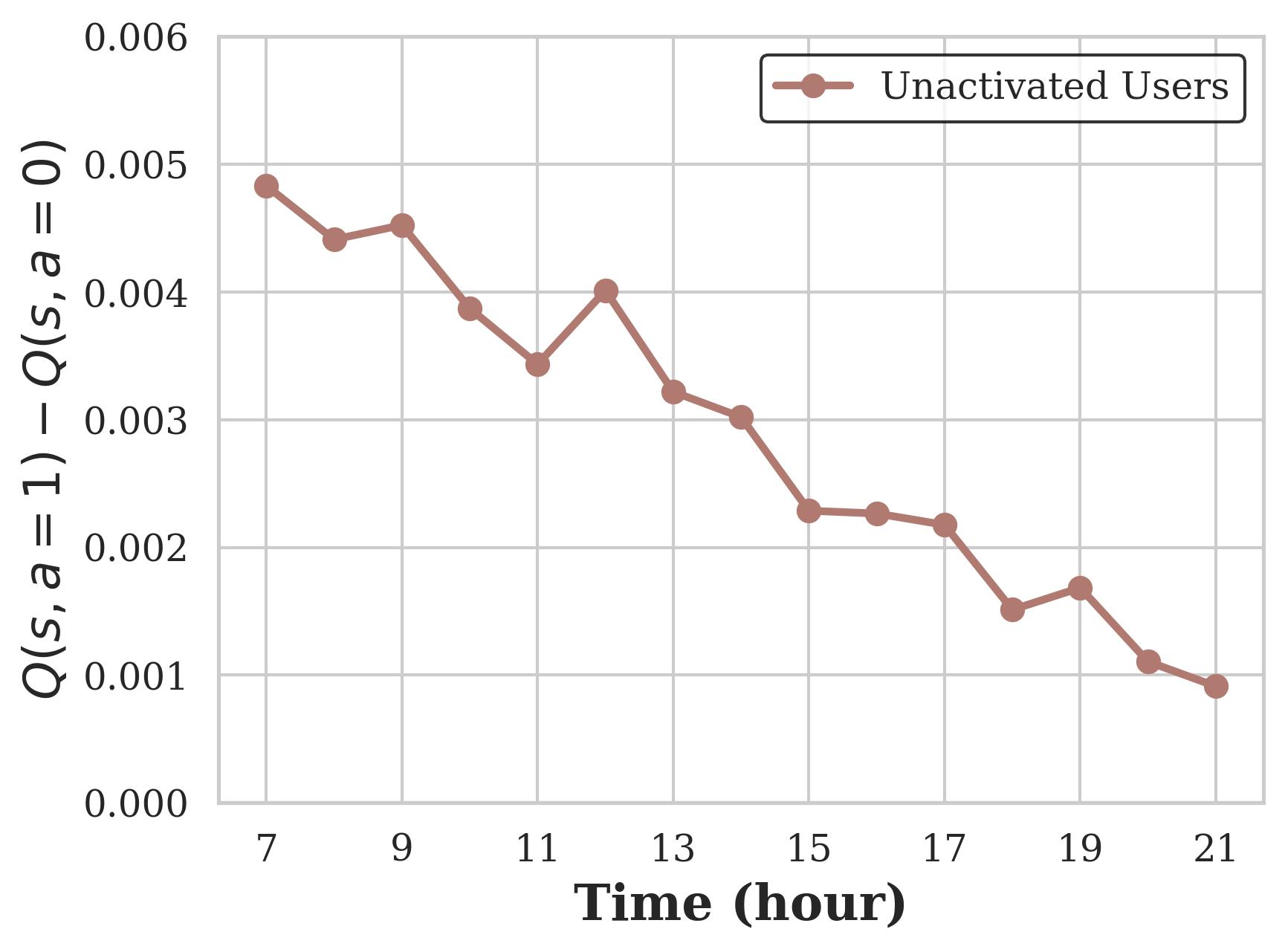}
        \caption{Action value difference for unactivated users.}
        \label{fig:q_diff}
    \end{subfigure}
    % 可选：添加整个大图的总标题（学术论文通常需要）
    \caption{Analysis of the learned $Q$ value.}
    \label{fig:q_value_analysis}
\end{figure}

% \noindent\textbf{Alleviation of Extreme Sending Behaviors.} In addition to optimizing long-term retention, Bellman RTG based modeling can also significantly improve the immediate user experience by naturally restricting irrational sending decisions. As detailed in Table~\ref{tab:send_distribution}, the distribution of daily sends per user becomes substantially more balanced under our proposed method. The system effectively curtails extreme over-sending, reducing the number of users receiving more than $20$ and $30$ pushes per day by 6.87\% and 9.30\%, respectively. Simultaneously, the volume of zero-send cases (users receiving no pushes at all) decreases by 4.06\%. This confirms that modeling Bellman based RTG penalizes sending too frequently or too infrequently and smoothly shifting the overall population towards a more reasonable and moderate daily push volume (1$\sim$20 sends).
\noindent\textbf{Alleviation of Extreme Sending Behaviors.} In addition to optimizing long-term retention, the Bellman RTG formulation significantly improves immediate user experience by naturally restricting suboptimal sending decisions. As detailed in Table~\ref{tab:send_distribution}, the distribution of daily sends per user becomes substantially more balanced. The system effectively curtails excessive sending, reducing the proportion of users receiving more than 20 and 30 pushes per day by 6.87\% and 9.30\%, respectively. Simultaneously, zero-send cases (users receiving no pushes at all) decrease by 4.06\%. This indicates that modeling the Bellman-based RTG effectively penalizes both overly frequent and overly sparse sending, smoothly shifting the overall population toward a more reasonable and moderate daily push volume (1$\sim$20 sends).

% \noindent\textbf{Accuracy of the Learned $Q$ Value.}  Taking $Q^+$ as an example, we conduct further analysis to verify the accuracy of $Q$ value in the execution agent, as shown in Figure~\ref{fig:q_calibration} and Figure~\ref{fig:q_diff}. First, the ratio of predicted Q value to actual ground truth RTG remains highly stable between 1.0 and 1.1 across all hours of the day, demonstrating accurate value predictions. Second, we investigate the marginal value of sending a push by analyzing 
% $Q^+(s,a=1)-Q^+(s,a=0)$. For unactivated users, this difference exhibits a monotonic trend, matching the business intuition that users are much easier to become active earlier in the day. This observation is also consistently supported by the temporal dynamics presented earlier in Figure~\ref{fig:time}.
\noindent\textbf{Accuracy of the Learned $Q$-Values.} Taking $Q^+$ as an example, we perform a further analysis to evaluate the accuracy of the value estimation of the execution agent, as shown in Figure~\ref{fig:q_calibration} and Figure~\ref{fig:q_diff}. First, the ratio of the predicted $Q$-value to the actual ground-truth RTG remains highly stable between 1.0 and 1.1 across all hours of the day, demonstrating well-calibrated predictions. Second, we investigate the marginal value of triggering a push by analyzing $Q^+(s,a=1)-Q^+(s,a=0)$. For currently unactivated users, this marginal difference exhibits a decreasing monotonic trend throughout the day. This effectively captures the underlying population dynamics: as time advances, the remaining pool of unactivated users becomes increasingly dominated by inherently low-engagement users, naturally leading to a diminished marginal return for late-day pushes. This observation is also consistently supported by the temporal dynamics presented earlier in Figure~\ref{fig:time}.

\begin{table}[t]
\centering
\caption{Relative computational overhead across agents (normalized to the filtering agent).}
\begin{tabular}{lrr}
\toprule
& \multicolumn{2}{c}{\textbf{Resource}} \\
\cmidrule(lr){2-3}
Agent & Overall & Per 1k requests \\
\midrule
Filtering & 1.00 & 1.00 \\
Planning & 1.27 & 1.27 \\
Execution & \textbf{9.62} & \textbf{57.72} \\
\bottomrule
\end{tabular}
\label{tab:service_resource_consumption}
\end{table}

\subsection{Further Analysis on Filtering}
% 加入filter的意义
% \noindent\textbf{Necessity of Early-Stage Filtering.} We first compare the resource consumption across different agent's to demonstrate the necessity of early-stage filtering for real-world online deployment. As shown in Table~\ref{tab:service_resource_consumption}, although the filtering and planning agents share a comparable per-request overhead, the execution agent is computationally far more expensive. Specifically, it consumes over 50 times the computing resources of the filtering stage per 1k requests. This massive overhead is primarily driven by the complex candidate item sorting mechanism within this agent, which requires intensive, real-time value estimation for multiple candidate items. Consequently, effectively reducing the volume of requests that reach this final execution stage translates into substantial computational savings, ensuring that the end-to-end agentic pipeline remains highly scalable and viable under massive online traffic.
\noindent\textbf{Necessity of Early-Stage Filtering.} We first compare the resource consumption across different modules to demonstrate the necessity of early-stage filtering for real-world online deployment. As shown in Table~\ref{tab:service_resource_consumption}, although the filtering and planning agents share a comparable per-request overhead, the execution agent is computationally far more expensive. Specifically, it consumes over 50 times the computing resources of the filtering stage per 1k requests. This massive overhead is primarily driven by the complex candidate item sorting mechanism within this agent, which requires intensive, real-time value estimation for multiple candidate items. Consequently, effectively reducing the volume of requests that reach this final execution stage translates into substantial computational savings, ensuring that the end-to-end agentic pipeline remains highly scalable and viable under massive online traffic.

\begin{figure}[t]
    \centering
    \Description{Online long-term value change and item-sorting QPS reduction under minimum trigger gaps from zero to twenty-four hours. QPS reduction grows with the trigger gap, while aggressive hour-scale filtering substantially harms long-term value.}
    \includegraphics[width=0.75\linewidth]{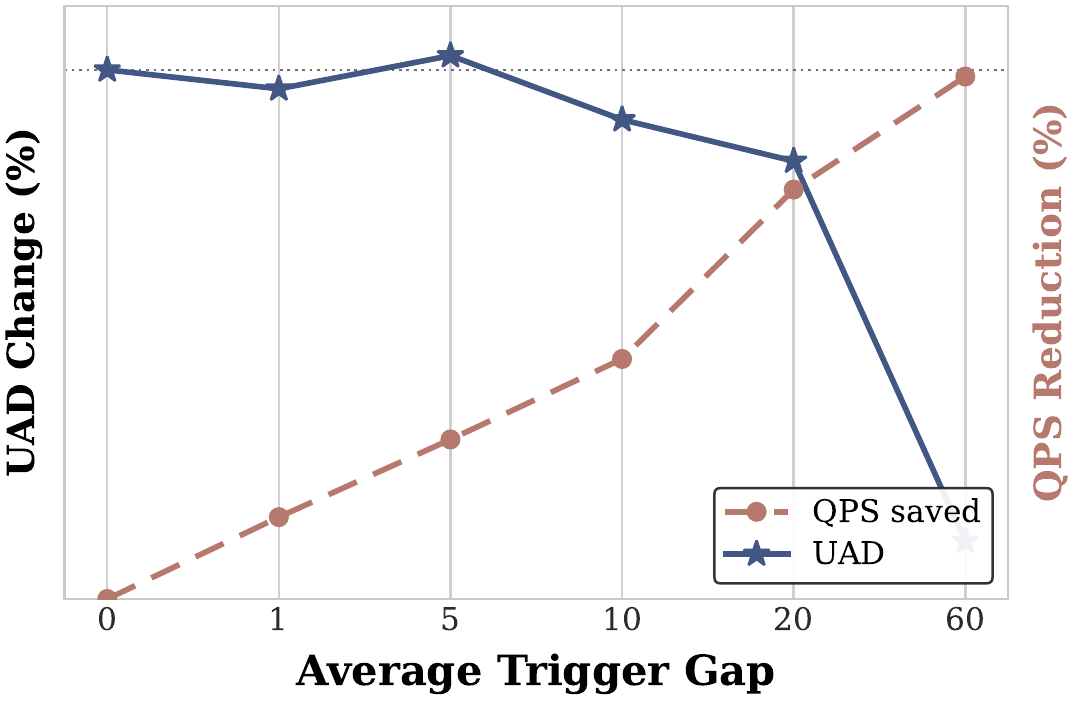}
    \caption{Online sensitivity to the fixed filtering threshold. The two axes report UAD and QPS reduction, respectively.}
    \label{fig:filtering_threshold_sensitivity}
\end{figure}

\noindent\textbf{Sensitivity of the Filtering Threshold.} In the filtering agent, we use a fixed threshold to filter out low-value events generated by the planning agent or real-time business (e.g., friends posting new videos). Adjusting this threshold affects the triggering frequency of the execution agent. We tune the threshold and monitor the time interval between two triggers, termed the \textit{trigger gap}, as a key metric in our production environment. Figure~\ref{fig:filtering_threshold_sensitivity} reveals a clear inflection point in the value-efficiency trade-off. As the average trigger gap increases by up to 20 minutes, QPS savings increase steadily and significantly, while UAD experiences only a marginal decline. However, more aggressive filtering incurs a disproportionate value loss: the UAD drops sharply when the average trigger gap reaches 60 minutes. We therefore adopt a conservative minute-scale threshold rather than an hour-scale rule to optimally balance efficiency and user value.

% \begin{table}[t]
% \centering
% \caption{Request drop compared with no filter and 20\% random filter. We present overall request and the results under different activity levels.}
% \resizebox{\linewidth}{!}{\begin{tabular}{lccc}
% \toprule
% & \multicolumn{3}{c}{\textbf{Activity Level}} \\
% \cmidrule(lr){2-4}
% Compared Scheme & \textbf{Inactive} & \textbf{High-Active} & \textbf{Fully-Active} \\
% \midrule
% No Filter  & -78.86\% & -66.63\% & -85.65\% \\
% Random Filter & -5.600\% & -5.131\% & -63.70\% \\
% \bottomrule
% \end{tabular}}
% \label{tab:resource_performance}
% \end{table}
\begin{table}[t]
\centering
\caption{Request drop rates across different user activity levels. Values represent the percentage of requests pruned by our filtering agent compared to the No-Filter and Random-Filter baselines.}
\resizebox{\linewidth}{!}{\begin{tabular}{lccc}
\toprule
& \multicolumn{3}{c}{\textbf{Activity Level}} \\
\cmidrule(lr){2-4}
\textbf{Baseline} & \textbf{Inactive} & \textbf{High-Active} & \textbf{Fully-Active} \\
\midrule
No Filter & 78.86\% & 66.63\% & 85.65\% \\
Random Filter & 5.60\% & 5.13\% & 63.70\% \\
\bottomrule
\end{tabular}}
\label{tab:resource_performance}
\end{table}

\noindent\textbf{Value-Aware Resource Reduction.} Table~\ref{tab:resource_performance} further demonstrates that the filtering agent provides substantial resource reduction by aggressively pruning low-value requests before they enter the more expensive downstream agents. Compared to the no-filter setting, the proposed filtering agent achieves an overall request drop rate of 74.88\%, with particularly high reduction rates for inactive (78.86\%) and fully-active (85.65\%) users. This aligns perfectly with our design motivation. For inactive users, the marginal response value is inherently low, making them prime candidates for early pruning to save compute. Conversely, fully-active users naturally trigger an abundance of business events (such as followed creators live starts or friends posting new videos), generating numerous potential push opportunities. Processing and sending all these redundant pushes yields negligible value while significantly harming the user experience by causing notification fatigue.

Consequently, the filtering module predominantly targets these redundant requests. To achieve the exact same final push send volume, a naive random filter must allow significantly more requests to pass into the downstream pipeline, wasting compute on low-quality opportunities that the execution agent will ultimately reject. In contrast, our value-aware filtering agent accurately predicts and prunes these doomed requests upfront. Compared to this random baseline, our agent removes far more requests overall, specifically avoiding wasted computation on fully-active users (achieving an additional 63.70\% request drop rate on this group). This explicitly demonstrates that its pruning decisions are driven by user value rather than merely a fixed traffic budget.

These results further reveal that the filtering stage plays two distinct roles in the system. (i) By discarding a large portion of candidate push events early, it directly slashes the computational cost of the subsequent pipeline. (ii) By selectively pruning redundant and low-value push opportunities, it safeguards against over-messaging, significantly reducing NE without compromising UAD.

% \begin{table}[t]
% \centering
% \caption{Effect of varying active send gap on User, negative experience, and resource consumption.}
% \begin{tabular}{c|c|c|c}
% \toprule
% Active send gap & UAD$\uparrow$ & NE$\downarrow$ & Resource$\downarrow$ \\
% \midrule
% 4h  & +0.0224\% & -0.0059\% & +0.0312 \\
% 8h  & +0.0381\% & +0.0033\% & -1.462 \\
% 12h & +0.0190\%  & -0.0066\% & -2.897 \\
% 16h & +0.0136\% & -0.0001\% & -2.000 \\
% 20h & -0.0260\%  & -0.0135\% & -3.896 \\
% 24h & -0.1529\% & -0.0105\% & -4.538 \\
% \bottomrule
% \end{tabular}
% \label{tab:active-send-gap}
% \end{table}

% \begin{table}[t]
% \centering
% \caption{Effect of varying send gap on user activity, negative experience, and resource consumption.}
% \begin{tabular}{c|c|c|c}
% \toprule
% Send gap & UAD$\uparrow$ & NE$\downarrow$ & Resource$\downarrow$ \\
% \midrule
% 10min  & +0.0263\% & -0.0053\% & +2.699 \\
% 20min  & +0.0395\% & -0.0065\% & +0.681 \\
% 40min  & -0.0185\% & -0.0119\% & -1.073 \\
% 60min  & -0.0888\% & -0.0007\% & -4.348 \\
% 120min & -0.2791\% & -0.0068\% & -10.693 \\
% \bottomrule
% \end{tabular}
% \label{tab:send-gap}
% \end{table}

\section{Conclusion}

In this paper, we addressed the highly non-trivial "whether and when" delivery problem in large-scale push recommendation systems. Existing solutions, constrained by pre-planned frequencies or fixed-interval triggering, inherently operate as passive systems that struggle with rigid schedules and local optima. To overcome these limitations, we proposed STEPS, a proactive, self-triggered end-to-end agentic push recommendation system. STEPS reformulates push recommendation as a self-triggered agentic process in which the system decides not only whether to act, but also when to invoke itself again, thereby operating in a closed loop. Specifically, STEPS utilizes two decision transformer-based agents: a planning agent employing a novel gated ordinal regression method to schedule the next system invocation, and a value-guided execution agent that decides whether to send the push based on trajectory rewards. Additionally, we incorporated a lightweight filtering agent to screen out low-value requests before they consume heavy computation and to safeguard against unreasonable planning behaviors. Extensive online A/B testing on Douyin, a platform with over 1 billion users, demonstrated that STEPS significantly improves long-term re-engagement by increasing user active days by \textbf{0.2843\%} while decreasing the push permission disablement rate by \textbf{1.9089\%}. Furthermore, the filtering agent successfully reduced computational overhead by \textbf{79.42\%}.

\begin{acks}
Z. Lin is supported by the Beijing Major Science and Technology Project (No. Z251100008425006), the Beijing Natural Science Foundation (No. L257007), the NSF China (No. 62276004), and the State Key Laboratory of General Artificial Intelligence.
\end{acks}

\newpage
\bibliographystyle{ACM-Reference-Format}
\bibliography{sample-base}

\end{document}